\documentclass[a4paper,papersize]{eptcs}

\usepackage{etex}
\usepackage[utf8]{inputenc}
\usepackage[english]{babel}
\usepackage{amsmath}
\usepackage{amsthm}
\usepackage{amssymb}
\usepackage{amsfonts}
\usepackage{mathtools}
\usepackage{stmaryrd} 
\usepackage{listings}
\usepackage[T1]{fontenc}

\usepackage{wrapfig}

\usepackage{aliascnt}
\usepackage{hyperref}

\usepackage{multirow}

\usepackage[retainorgcmds]{IEEEtrantools} 

\newcommand{\code}[1]{\texttt{\textsl{{#1}}}}

\newcommand{\dotcup}{\ensuremath{\mathaccent\cdot\cup}}

\newcommand{\xhra}{\xhookrightarrow}

\newcommand{\ASet}{A}
\newcommand{\USet}{U}
\newcommand{\GSet}{G}

\newcommand{\taarrow}{\rightarrow}

\renewcommand{\dotcup}{\mathbin{\mathaccent\cdot\cup}}

\DeclareMathOperator{\nil}{NIL}
\DeclareMathOperator{\base}{BASE}
\DeclareMathOperator{\callr}{RCALL}
\DeclareMathOperator{\calln}{NCALL}
\DeclareMathOperator{\acq}{ACQ}
\DeclareMathOperator{\ret}{RET}
\DeclareMathOperator{\use}{USE}
\DeclareMathOperator{\spawn}{SPAWN}
\DeclareMathOperator{\cut}{CUT}

\DeclareMathOperator{\mif}{\mathbf{if}}

\DeclareMathOperator{\sptop}{\text{\small s} \hspace{-1.3ex} {\top}}
\DeclareMathOperator{\spbot}{\text{\small s} \hspace{-1.3ex} {\bot}}

\DeclareMathOperator{\Act}{Act}

\DeclareMathOperator{\post}{post}
\DeclareMathOperator{\lspost}{lspost}
\DeclareMathOperator{\conf}{conf}

\usepackage{tikz}
\usetikzlibrary{%
  arrows,%
  shapes.misc,
  shapes.arrows,%
  chains,%
  matrix,%
  positioning,
  scopes,%
  decorations.pathmorphing,
  shadows%
}

\usetikzlibrary{shapes.multipart,shapes.symbols} 
\usetikzlibrary{mindmap,trees}
\usetikzlibrary{plothandlers}
\usetikzlibrary{automata}
\usetikzlibrary{backgrounds}

\tikzset{
  nonterminal/.style={
    rectangle,
    minimum size=6mm,
    very thick,
    draw=red!50!black!50,         
    top color=white,              
    bottom color=red!50!black!20, 
    font=\itshape
  },
  terminal/.style={
    rounded rectangle,
    minimum size=6mm,
    very thick,draw=black!50,
    top color=white,bottom color=black!20,
    font=\ttfamily},
  skip loop/.style={to path={-- ++(0,#1) -| (\tikztotarget)}}
}

{
  \tikzset{terminal/.append style={text height=1.5ex,text depth=.25ex}}
  \tikzset{nonterminal/.append style={text height=1.5ex,text depth=.25ex}}
}

\newsavebox{\fmbox}

\lstdefinelanguage{xsb}{%
  language=Prolog,%
  morekeywords={table , assert , write , nl},%
  moredelim=*[s][\itshape]{:-}{.},%
  moredelim=**[is][\ttfamily]{|}{|},%
  keywordstyle={\bf},%
  basicstyle={\ttfamily}%
}%
\lstdefinelanguage{MyJava}{%
  language=Java,%
  morekeywords={start , print, sync},%
  keywordstyle={\bf}%
}%
\addto\extrasenglish{%
}

\theoremstyle{plain}

\newaliascnt{lem}{thm} 

\aliascntresetthe{lem}

\newaliascnt{proposition}{thm}
\newtheorem{proposition}[proposition]{Proposition}
\aliascntresetthe{proposition}

\newaliascnt{cor}{thm}  

\aliascntresetthe{cor}

\theoremstyle{definition}
\newtheorem{definition}{Definition}

\newtheorem{example}{Example}

\theoremstyle{remark}

\hypersetup{
    colorlinks,
    citecolor=black,
    filecolor=black,
    linkcolor=black,
    urlcolor=black
}

\newlength{\rulenegskip}
 \title{Iterable Forward Reachability Analysis of Monitor-DPNs}

 \author{Benedikt Nordhoff \qquad \qquad Markus
   Müller-Olm \institute{Institut für Informatik, Westfälische Wilhelms-Universität
   Münster\footnote{This work was partially funded by DFG projects IFC
     for Mobile Components (MU 1508/2) within priority program RS3
     (SPP
     1496), and OpiAT (MU 1508/1).}}
   \email{\quad\quad\qquad b.n@wwu.de \qquad\qquad\qquad markus.mueller-olm@wwu.de}
   \and
   Peter Lammich
   \institute{Fakultät für Informatik, TU München}
   \email{lammich@in.tum.de}}
 
\newcommand{\authorrunning}{Nordhoff, Müller-Olm, Lammich}
\newcommand{\titlerunning}{Iterable Forward Reachability Analysis of Monitor-DPNs}

\begin{document}
\renewcommand\bibname{References}
\maketitle

\begin{abstract}
  There is a close connection between data-flow analysis and model
  checking as observed and studied in the nineties by Steffen and
  Schmidt.  This indicates that automata-based analysis techniques
  developed in the realm of infinite-state model checking can be
  applied as data-flow analyzers that interpret complex control
  structures, which motivates the development of such analysis
  techniques for ever more complex models.  One approach proposed by
  Esparza and Knoop is based on computation of predecessor or
  successor sets for sets of automata configurations.  Our goal is to
  adapt and exploit this approach for analysis of multi-threaded Java
  programs.  Specifically, we consider the model of Monitor-DPNs for
  concurrent programs.  Monitor-DPNs precisely model unbounded
  recursion, dynamic thread creation, and synchronization via
  well-nested locks with finite abstractions of procedure- and
  thread-local state.  Previous work on this model showed how to
  compute regular predecessor sets of regular configurations and
  tree-regular successor sets of a fixed initial configuration.  By
  combining and extending different previously developed techniques we
  show how to compute tree-regular successor sets of tree-regular
  sets.  Thereby we obtain an iterable, lock-sensitive forward
  reachability analysis.  We implemented the analysis for Java
  programs and applied it to information flow control and data race
  detection.
\end{abstract}

\section{Introduction}
\label{sec:introduction}

Developing, debugging and analysing parallel programs is notoriously
hard because the actual scheduling of different threads is mostly
unspecified.  For the developer it requires a lot of experience and
skill to foresee all possible relevant impacts of the unknown
scheduling.  Debugging is difficult as critical bugs may only manifest
under scheduling conditions that are hard to provoke during testing,
but might be typical in a particular environment.  Moreover,
scheduling-dependent bugs are often hard to reproduce.  From the
program analysis side, parallel programs are tough as one quickly
reaches the decidability barrier.  Of particular relevance is the
result by Ramalingam \cite{Ramalingam:2000:CSA:349214.349241} that
full context and synchronization sensitivity in the presence of
rendezvous-style synchronization and recursion renders even
reachability undecidable.  Kahlon, Ivancic and Gupta
\cite{DBLP:conf/cav/KahlonIG05} demonstrated how rendezvous-style
synchronization can be modeled by non-well-nested locks.  On another
track M\"uller-Olm and Seidl \cite{Muller-Olm:2001:OSP:380752.380864}
showed that optimal slicing for recursive parallel programs is
undecidable. The goal therefore is to find reasonable abstractions
where the problem of interest is effectively decidable, but sufficient
precision is retained.

Steffen observed in the early nineties that data-flow analysis
problems can be solved via model checking \cite{steffen-dfa-mc}, an
observation elaborated later by Schmidt \cite{schmidt-dfa-mc-ai} and
Schmidt and Steffen \cite{schmidt-steffen}. This observation indicates
that automata-based analysis procedures developed in the realm of
infinite-state model checking can be applied as data-flow analyzers
that interpret complex control structures. One approach proposed by
Esparza and Knoop \cite{Esparza:1999} is based on computation of
predecessor or successor sets for sets of automata configurations.
Our goal is to adapt and exploit this approach for analysis of
multi-threaded Java programs.  More specifically, we utilize an
automata-theoretic approach which handles full recursion, thread
creation and synchronization via reentrant well-nested locks.  The
underlying model is that of Dynamic Pushdown Networks (DPN) which was
introduced (without locks) by Bouajjani {\it et al.}
\cite{concur05-rsa-dpn} in 2005.  They showed how the $pre^{*}$
operator for calculating predecessor sets of configurations -- words
over an alphabet of control and stack symbols -- preserves regularity
in the same way as done earlier in the sequential case
\cite{Bouajjani:1997:RAP:646732.701281}.  Later, the DPN-model was
enriched to allow for scheduling restrictions based on well-nested
locks \cite{cav09-ps-dpn-trc}.  This was done by extending the {\em
  acquisition history} technique of Kahlon {\it et al.}
\cite{DBLP:conf/cav/KahlonIG05} and thereby obtaining a tree-regular
characterization of lock-sensitive schedulability, which was encoded
into the control states of the DPN.  The converse set of
configurations reachable from a given configuration ($post^{*}$) was
shown to be non-regular in the word semantics \cite{concur05-rsa-dpn}.
Gawlitza {\it et al.} \cite{join-lock-sens-fwd} then switched from
configuration words to \emph{execution trees}.  An execution tree
describes the steps of an execution as well as the reached
configuration.  The tree structure makes visible both the parallel
execution of steps in different threads and the nesting structure of
procedure calls and returns.  This allowed them to obtain a
tree-regular representation of the set of execution trees reachable
from an initial single thread configuration.  Note that the tree-based
semantics is strictly more expressive than the word-based semantics:
By a suitable traversal of the execution tree one can obtain the
configuration word and every regular property over configuration words
can be translated to an equivalent tree-regular property over
execution trees.  Conversely, the execution tree cannot be obtained
from the configuration word as it contains information of the history
about the execution. The applicability of these techniques for the
analysis of programs in real programming languages like Java was
postulated by different authors but never implemented.

We extend this research by the following three main contributions:
\begin{enumerate}
\item In a first step we adapt the forward analysis approach of
  Gawlitza {\it et al.}  \cite{join-lock-sens-fwd} to fit the setting
  of well-nested reentrant locks as found in Java.  The resulting
  technique can treat reachability problems like checking a program
  for data races.
\item In order to solve data-flow problems, like calculating def-use
  dependencies, we extend our technique to calculate $post^{*}$ for arbitrary
  tree-regular sets of reachable
  configurations.  The extension is based on the insight that each
  execution tree that is reached in the course of an execution can be
  obtained from any later reached tree.  In order to check for lock
  sensitive schedulability we adapt techniques based on so
  called \emph{acquisition} and \emph{release structures}
  \cite{DBLP:conf/lics/KahlonG06,cav09-ps-dpn-trc}.  We do this by
  identifying previously defined criteria for schedulability and
  checking those in a modular way by defining appropriate tree
  automata.
\item We report on our implementation of the developed techniques for
  the analysis of Java programs.  Our application includes an Eclipse
  IDE plugin for data race detection and an inter-thread def-use
  dependency checker for an information flow control analysis tool.
\end{enumerate}
 
As we head for an actual implementation, we describe all constructions
more explicitly than done in previous work.  We will explain the key
insights from the original correctness proofs of the used techniques.
Afterwards, we report on obstacles, solutions and results from our
implementation for the analysis of parallel Java programs.  We also
provide an undecidability result in the presence of \code{wait}-calls
as found in Java.

\subsection{Related Work / History}
\label{sec:related-work}
This work can be related to several lines of previous research.  There
is a vast amount of literature on analyzing infinite-state systems
with automata-based techniques. We only cover a few papers here that
are closely related to our work.  Dating back till 1964 is research
concerning the preservation of regularity under various term rewrite
systems. Büchi's work on \emph{Regular Canonical Systems}
\cite{buechi-regular-systems}, which, among other things, implies that
the set of reachable configurations of a push-down automata is
regular, is probably one of the first. In 1997, Bouajjani, Esparza and
Maler \cite{Bouajjani:1997:RAP:646732.701281} showed how predecessor
sets of regular sets of push-down automata configurations are
effectively regular. They used their results to perform model checking
of linear and branching time logic of push-down automata.  This work
was the basis for the transition to multi-threaded programs in the
previously mentioned work by Bouajjani {\it et al.}
\cite{concur05-rsa-dpn} from 2005 where Dynamic Pushdown Networks were
introduced.  Lugiez and Schnoebelen in 1998
\cite{lugiez-schnoebelen-pa-concur, lugiez-schnoebelen-regular-pa}
applied tree automata-based analysis techniques to another class of
parallel systems, so called PA-processes and showed that
tree-regularity of sets of PA-process terms is preserved by pre$^{*}$
and post$^{*}$ operations.  Then in 2000 Esparza and Podelski
\cite{esparza-podelski-pre-post-ipfg} showed how these sets can
effectively be calculated as the least model of a set of Horn-clauses,
similar to the encoding in logical programs that we use in our
implementation.  This could then be used to solve bitvector data-flow
analysis problems.  Also in 2000, Bouajjani {\it et al.}
\cite{bouajjani-et-al-context-free-grammars} interpreted context
free-grammars as term rewriting systems over regular sets of words and
gave an efficient algorithm to compute pre$^{*}$ images. This could be
used to efficiently answer various problems concerning context
free-grammars.  They also stated an equivalent encoding as a set of
Horn-clauses which led to an algorithm of equal time complexity.

Work on how to use automata theoretic and model checking approaches to
solve data-flow analysis problems includes work by Steffen
\cite{steffen-dfa-mc} from 1991 who showed how data-flow analysis
problems can be expressed as model checking problems of modal
mu-calculus formulas against a finite program model.  Further research
in this direction was done in 1998 by Schmidt partially in
collaboration with Steffen \cite{schmidt-dfa-mc-ai,schmidt-steffen}
who incorporated abstract interpretation in this approach. Work on
solving data-flow analysis problems via pre$^{*}$ and post$^{*}$
computations of regular sets was done by Esparza and Knoop in 1999
\cite{Esparza:1999}.

\subsection{Motivation}
\label{sec:motivation}

This work was motivated by the goal to improve the precision of an
information flow control analysis for parallel Java programs based on
system dependence graphs (SDG) as described for example by Hammer and
Snelting \cite{hammer09ijis} or Giffhorn \cite{giffhorn12thesis}.  The
specific aim was to improve the handling of data flows between
different threads in the SDG-based analysis. Besides, we obtain the
first implementation of the previously developed techniques for real
world applications.

In order to illustrate the effects of locking and thread creation,
consider the snippets of Java code in \autoref{tab:ex1}.  Assuming an
adequate context where \lstinline|x| and \lstinline|y| are initially 0
and \lstinline|t2| is an object of the Java class Thread: Can the
programs print the value 42? The answer is no for all programs, but
how can an analysis automatically infer this? An analysis abstracting
from thread creation must assume that 42 can be printed by the first
program.  The second example shows the interplay between thread
creation and locking.  The main thread holds a lock while spawning the
second thread and only prints the variable before releasing the lock.
However, the second thread needs to use the lock before writing 42 to
the variable.  In the third example the write of 42 would need to be
scheduled between the assignment \lstinline|x = 17| and the print
which is imposible due to the shared lock.  The fourth example is
similar from a scheduling point of view but there are two transfers
involved.  While both transfers are feasible, there is no run
exhibiting both.  In the fifth example, analogously to the third
example, the assignment of 42 to $x$ must be scheduled right before
printing $x$, but there is no shared lock held by both processes at
these points. By exhaustive inspection of the different possibilities
in which the two threads can acquire and release the locks $a$ and $b$
one can see that it is not possible to schedule the assignment as
required. The last example is similar but a little more involved.
Either there is an intervening kill or the program reaches a deadlock.
A lock-sensitive DPN-based analysis as described in the remainder of
this paper can treat all these effects precisely and automatically
infers that none of the example programs can print the value 42.%

\begin{table}[t]
  \small
  \centering
  \begin{tabular}{ll@{\hspace{2em}}l}  
    \#&main method&\lstinline|t2|'s run method\\\hline
1&
\begin{lstlisting}
print(x);
t2.start();
\end{lstlisting}  &  
\begin{lstlisting}
x = 42;
\end{lstlisting}\\\hline
  2&
\begin{lstlisting}
synchronized(a){
  t2.start();
  print(x);    }
\end{lstlisting}  &
\begin{lstlisting}
synchronized(a){$\dots$}
x = 42;
\end{lstlisting}\\\hline
  3&
\begin{lstlisting}
t2.start();
synchronized(a){
  x = 17;
  print(x);    }
\end{lstlisting}  &
\begin{lstlisting}
synchronized(a){
  x = 42;
}
\end{lstlisting}\\\hline
  4&         
\begin{lstlisting}
t2.start();
synchronized(a){
  y = 42;
  print(x);    }
\end{lstlisting} &
\begin{lstlisting}
synchronized(a){
  x = y;
}
\end{lstlisting} \\\hline
  5&         
\begin{lstlisting}
t2.start();
synchronized(a){  
  synchronized(b) {$\dots$}
  x = 17;
  print(x);    }
\end{lstlisting} &
\begin{lstlisting}
synchronized(b){
  synchronized(a) {$\dots$}
  x = 42;
}
\end{lstlisting} \\\hline
  6&
\begin{lstlisting}
t2.start()
synchronized(a){
  synchronized(b){
    x = 42;      }  
  x = 23;      }
\end{lstlisting}&
\begin{lstlisting}
synchronized(b){
  if ($\dots$) { 
    synchronized(a){$\dots$} 
  } else    { 
    x = 17; }
  print(x);    }
\end{lstlisting}
  \end{tabular}%
    \caption[ ]{Spurious examples of data flows between threads}%
  \label{tab:ex1}%
\end{table}%


\section{The Monitor-DPN Model}
\label{sec:monitor-dpn-model}
We now define the underlying model for our analyses.  DPNs precisely
model unbounded recursion and thread creation with finite abstractions
of method- and thread-local state but abstract from global state.
Intuitively, a DPN is a set of push-down systems which are able to add
new push-down systems as a side effect of their transitions.  The
latter corresponds to thread creation.  In a Monitor-DPN we also allow
the threads to communicate via a finite set of reentrant locks, which
are used in a well-nested fashion.  Reentrance means that a thread may
acquire the same lock multiple times and releases it only after a
matching number of release-operations.  We enforce well-nestedness
syntactically by only allowing a lock to be acquired when pushing a
local state onto the stack and releasing it implicitly when the old
stack level is reached again.  That is, lock acquisition and release
is bound to procedure calls.  Note that if locks are used in a
syntactically well-nested fashion -- like in synchronized blocks in
Java -- they can be transformed to procedure calls with attached
locks.  We call locks used in this way \emph{monitors}.

\begin{definition}
\label{def:dpn}
A \emph{Monitor-DPN} $\mathcal{M}$ is a tuple
$(\Act,P,\Gamma,X,\Delta,(p_{0},\gamma_{0}))$ consisting of an initial
configuration $(p_{0},\gamma_{0}) \in P \times \Gamma$ and finite sets
of: actions $\Act$, control states $P$, stack symbols $\Gamma$, locks
$X$ and a set $\Delta$ of transition rules of the form:
  \begin{center}
    (Base)\ $p\gamma\xhra{a}p'\gamma'$ 
      \qquad (Call)\ $p\gamma\xhra{a}p'\gamma'\gamma_{r}$
      \qquad (Return)\ $p\gamma\xhra{a}p'$ \\*[1ex]
      (Spawn)\ $p\gamma\xhra{a}p_s\gamma_sp'\gamma'$
      \qquad(Monitor)\ $p\gamma\xhra{a,x}p'\gamma'\gamma_{r}$
    \end{center}
    where $p,p',p_s \in P$, $a \in \Act$, $x \in X$,
    $\gamma,\gamma',\gamma_{r},\gamma_s \in \Gamma$ .  
    An \emph{ordinary DPN}
    is a Monitor-DPN without monitor rules.
\end{definition}

For the rest of this paper, we fix a Monitor-DPN $\mathcal{M} =
(\Act,P,\Gamma,X,\Delta,(p_{0},\gamma_{0}))$ and refer to Monitor-DPNs
generally as DPNs.  We define two semantics for DPNs, a lock-sensitive
and a lock-insensitive one.  To distinguish the threads in a
configuration, we use thread identifiers consisting of sequences of
natural numbers, which represent the way a thread was created. We
denote by $\tau \in \mathbb{N}^{*}$ the empty sequence and by $t\,n$
the sequence $t$ extended by $n$.  Moreover, we extend the set of
locks by a special \emph{no lock} symbol to $\bar X = X \dotcup
\{\bullet\}$, where we write $\dotcup$ for disjoint union.  We assume
that the sets $P$, $\Gamma$, $X$, $\Delta$, and $\mathbb{N}^{*}$ are
pairwise disjoint.

\begin{definition}
\label{def:dpn-sem}
A thread configuration is a word from $P(\Gamma\bar X)^{*}$.  A DPN
configuration $c \in 2^{\mathbb{N}^{*}\times P(\Gamma\bar X)^{*}}$ is
a finite set of pairs of thread identifiers and thread
configurations. The initial configuration is
$\{(\tau,p_{0}\gamma_{0}\bullet)\}$.  The set of locks \emph{held} by
a DPN configuration $c$ is $l(c) = \{ x \in X \mid \exists t,w,w'.\
(t,wxw') \in c\}$ and the set of {\em active} threads is $c|_{1} = \{
t \in \mathbb{N}^{*} \mid \exists w.\ (t,w) \in c\}$.  The
lock-sensitive semantics of DPN is given by the following transition
rules:

\noindent
\begin{IEEEeqnarray*}{r/c/l}
    c \dotcup \{(t,p\gamma yw)\} &\xrightarrow{
      p\gamma\xhra{a}p'\gamma'
    }_{t} &c \dotcup \{(t,p'\gamma'yw)\}\\
    c \dotcup \left\{ (t,p\gamma yw)\right\} &\xrightarrow{
      p\gamma\xhra{a}p' 
    }_{t} &c\dotcup \left\{ (t,p'w) \right\}\\
    c \dotcup \left\{ (t,p\gamma yw)\right\} &\xrightarrow{
      p\gamma\xhra{a}p'\gamma'\gamma_{r} 
    }_{t} &c\dotcup \left\{ (t,p'\gamma'\bullet\gamma_{r}yw) \right\}\\
    c \dotcup \left\{ (t,p\gamma yw)\right\} &\xrightarrow{
      p\gamma\xhra{a}p_s\gamma_sp'\gamma' 
    }_{t} &c\dotcup \left\{ (t,p'\gamma'yw) \,,\,
      (t\,n,p_{s}\gamma_{s}\bullet) \right\} \quad t\,n\not\in c|_{1} \wedge (n = 1 \vee t\,(n-1) \in c|_{1}) \\
    c \dotcup \left\{ (t,p\gamma yw)\right\} &\xrightarrow{
      p\gamma\xhra{a,x}p'\gamma'\gamma_{r} 
    }_{t} &c\dotcup \left\{ (t,p\gamma'x\gamma_{r}yw) \right\} \quad \text{if } x \not\in l(c) \enspace .
  \end{IEEEeqnarray*}
  The lock-insensitive semantics for DPN is obtained by dropping the
  constraint $x \not\in l(c)$ in the last rule.  The set $\Pi(\mathcal
  M)$ of executions of a DPN consists of all sequences $\pi$ of the
  form $\pi = c_{0}\xrightarrow{\eta_{1}}_{t_{1}} c_{1} \dots
  \xrightarrow{\eta_{n}}_{t_{n}} c_{n}$ where $c_{0} =
  \{(\tau,p_{0}\gamma_{0}\bullet)\}$ and $\eta_{i} \in \Delta$.  We
  denote by $\delta(\pi) = \{\eta_{i} \mid 1\leq i \leq n\}$ the set
  of transitions used within execution $\pi$.
\end{definition}

For a set $C$ of DPN configurations and a set $\Delta' \subseteq
\Delta$ of transition rules, the set of configurations that are lock
sensitively reachable from $C$ via $\Delta'$ is
$\lspost_{\Delta'}^{*}(C) = \{c \mid \exists c_0\in C,\, \pi.\ \pi =
c_0\xrightarrow{\eta_{1}}_{t_{1}} c_{1} \dots
\xrightarrow{\eta_{n}}_{t_{n}} c \wedge \delta(\pi) \subseteq
\Delta'\}$. We also define the lock-insensitive version
$\post^{*}_{\Delta'}(C)$, which corresponds to the lock-insensitive
semantics.  The goal is to find a reasonable class of representations
for sets of configurations such that $\lspost^{*}_{\Delta'}$ is a
computable endomorphism on this class.

\subsection{Execution Trees}
\label{sec:execution-trees}

Note that for all configurations $c$ that are reached by executions
from the initial configuration,
the set of active threads $c|_{1}$ forms a tree and thread $t\,n$ is
the thread spawned by the n-th spawn transition performed by thread
$t$. If one denotes by $\pi^{t}$ the sequence of transitions performed
by $t$ in $\pi$ and builds a tree by making $\pi^{t\,n}$ the left
branch of the n-th spawn appearing in $\pi^{t}$ one obtains (up to
projection to the annotated actions) what is called an action tree
\cite{join-lock-sens-fwd,cav09-ps-dpn-trc}.  An action tree reflects
the constraints on the ordering of transitions to form a valid
execution: A transition must be executed after its predecessors in the
tree.  In the lock-insensitive setting, these constraints completely
characterize the valid executions, i.e., all topological orderings of
the transitions according to the action tree are valid executions.  In
the lock-sensitive setting, there are additional constraints imposed
by the locks, which are explained in
Section~\ref{sec:lock-sens-analys}.

As the $\pi^{t}$ correspond to traces of ordinary push-down systems,
they are essentially equivalent to context free languages.  One can
regain (tree-)regularity by adding additional structure which makes
the push-down behavior visible.  This is done by matching return
transitions to corresponding calls in the following way: First, we add
for convenience a special node $p\gamma$ to the end of $\pi^{t}$, if
the thread configuration reached by $t$ has a non-empty stack, where
$p$ is the reached control state and $\gamma$ is the top of stack.
Then let $\pi^{t} =
\eta_{0}\dots\eta_{i}\eta_{i+1}\dots\eta_{j}\eta_{j+1}\dots\eta_{k}$
(here $\eta_{k}$ may be the newly introduced node) and let $\eta_{i}$
be the first call- or monitor-rule in $\pi^{t}$ matched by a
return-rule, say $\eta_{j}$.  We transform $\pi^{t}$ to the tree
$\eta_{0}\dots\eta_{i}(\, \eta_{i+1}\dots\eta_{j}\, ,
\,\eta_{j+1}\dots\eta_{k}\, )$ and continue recursively with the new
subtraces $\eta_{i+1}\dots\eta_{j}$ and $\eta_{j+1}\dots\eta_{k}$.
Here we represent trees by terms.  If we again hook in the obtained
trees as left branches of the corresponding spawn operations, we
obtain the execution trees introduced by Gawlitza {\it et al.}
\cite{join-lock-sens-fwd}.  The execution tree of an execution $\pi$
is denoted by $\varcurlywedge(\pi)$.  Moreover, we define the
execution tree corresponding to the empty execution to be the initial
node $p_{0}\gamma_{0}$.

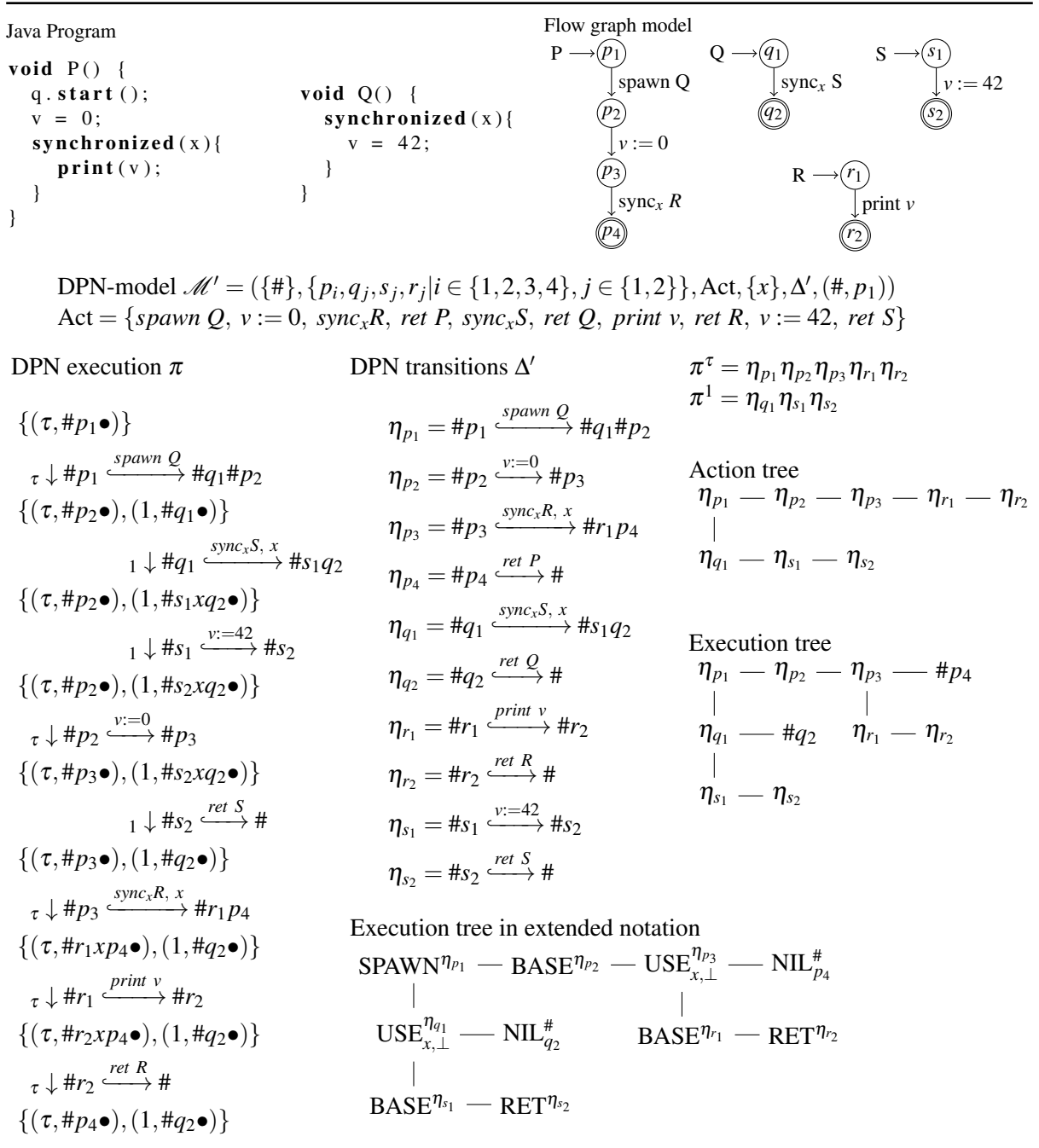
\begin{figure}
  \rule{\linewidth}{1pt}\\[1ex]
  \centering
  \begin{minipage}{.5\linewidth}
    \footnotesize
    Java Program\\
    \begin{minipage}{.45\linewidth}
\begin{lstlisting}
void P() {
  q.start();
  v = 0;
  synchronized(x){
    print(v);
  }
}
\end{lstlisting}
    \end{minipage}\hspace*{3em}%
    \begin{minipage}{.45\linewidth}
\begin{lstlisting}
void Q() {
  synchronized(x){
    v = 42;
  }
}
\end{lstlisting}
    \end{minipage}%
  \end{minipage}\hspace*{1em}%
  \begin{minipage}{.5\linewidth}
    \footnotesize
    Flow graph model\\
    \begin{tikzpicture}[initial text =P,node distance = 1.5em,shorten >= 1pt,
      every state/.style={minimum size=1.4em,inner sep=0pt}]
      \node[state,initial]      (p1)     {$p_{1}$};
      \node[state, below = of p1]      (p2)     {$p_{2}$};
      \node[state, below=of p2]      (p3)     {$p_{3}$};
      \node[state,accepting , below = of p3]      (p4)     {$p_{4}$};
      
      \path[->]  (p1) edge[]        node[right]       {spawn Q} (p2) 
      (p2)    edge[]        node[right]        {$v := 0$} (p3) 
      (p3)    edge[]        node[right]        {sync$_{x}\ R$} (p4) ;
      
      \begin{scope}[xshift=8em,initial text=Q]
        \node[state,initial]      (q1)     {$q_{1}$};
        \node[state,accepting , below = of q1]      (q2)     {$q_{2}$};        
        \path[->]  (q1) edge[]        node[right]       {sync$_{x}$ S} (q2) ;
      \end{scope}
      \begin{scope}[xshift=12em,yshift=-6em,initial text=R]
        \node[state,initial]      (r1)     {$r_{1}$};
        \node[state,accepting , below = of r1]      (r2)     {$r_{2}$};        
        \path[->]  (r1) edge[]        node[right]       {print $v$} (r2) ;
      \end{scope}
      \begin{scope}[xshift=16em,initial text=S]
        \node[state,initial]      (s1)     {$s_{1}$};
        \node[state,accepting , below = of s1]      (s2)     {$s_{2}$};        
        \path[->]  (s1) edge[]        node[right]       {$v := 42$} (s2) ;
      \end{scope}
    \end{tikzpicture}
  \end{minipage}

  \vspace{2ex}
  \begin{minipage}[t]{.9\linewidth}
    DPN-model $\mathcal{M'} = (\{\#\},\{p_{i},q_{j},s_{j},r_{j}|i \in
    \{1,2,3,4\}, j\in\{1,2\}\},\Act,\{x\},\Delta',(\#,p_{1}))$ \\
    $\Act = \{ spawn\ Q,\ v:=0,\ sync_{x}R,\ ret\ P,\ sync_{x}S,\
    ret\ Q,\ print\ v,\ ret\ R,\ v:=42,\ ret\ S \}$
  \end{minipage}

 \vspace{2ex}

  \begin{minipage}[t]{.33\linewidth}
    DPN execution $\pi$
    \begin{align*}
      \{&(\tau,\#p_{1}\bullet)\}\\
      &_{\tau}\downarrow \#p_{1} \xhra{spawn\ Q} \#q_{1}\#p_{2}\\
      \{&(\tau,\#p_{2}\bullet),(1,\#q_{1}\bullet)\}\\
      &\qquad\qquad_{1}\downarrow \#q_{1} \xhra{sync_{x}S,\ x} \#s_{1}q_{2}\\
      \{&(\tau,\#p_{2}\bullet),(1,\#s_{1}xq_{2}\bullet)\}\\
      &\qquad\qquad_{1}\downarrow \#s_{1} \xhra{v:=42} \#s_{2}\\
      \{&(\tau,\#p_{2}\bullet),(1,\#s_{2}xq_{2}\bullet)\}\\
      &_{\tau}\downarrow \#p_{2} \xhra{v:=0} \#p_{3}\\
      \{&(\tau,\#p_{3}\bullet),(1,\#s_{2}xq_{2}\bullet)\}\\
      &\qquad\qquad_{1}\downarrow \#s_{2} \xhra{ret\ S} \#\\
      \{&(\tau,\#p_{3}\bullet),(1,\#q_{2}\bullet)\}\\
      &_{\tau}\downarrow \#p_{3} \xhra{sync_{x}R,\ x} \#r_{1}p_{4}\\
      \{&(\tau,\#r_{1}xp_{4}\bullet),(1,\#q_{2}\bullet)\}\\
      &_{\tau}\downarrow \#r_{1} \xhra{print\ v} \#r_{2}\\
      \{&(\tau,\#r_{2}xp_{4}\bullet),(1,\#q_{2}\bullet)\}\\
      &_{\tau}\downarrow \#r_{2} \xhra{ret\ R} \#\\
      \{&(\tau,\#p_{4}\bullet),(1,\#q_{2}\bullet)\}      
    \end{align*}
  \end{minipage}%
  \begin{minipage}[t]{.33\linewidth}    
    DPN transitions $\Delta'$
    \begin{align*}
      &\eta_{p_{1}} = \#p_{1} \xhra{spawn\ Q} \#q_{1}\#p_{2}\\
      &\eta_{p_{2}} = \#p_{2} \xhra{v:=0} \#p_{3}\\
      &\eta_{p_{3}} = \#p_{3} \xhra{sync_{x}R,\ x} \#r_{1}p_{4}\\
      &\eta_{p_{4}} = \#p_{4} \xhra{ret\ P} \#\\
      &\eta_{q_{1}} = \#q_{1} \xhra{sync_{x}S,\ x} \#s_{1}q_{2}\\
      &\eta_{q_{2}} = \#q_{2} \xhra{ret\ Q} \#\\
      &\eta_{r_{1}} = \#r_{1} \xhra{print\ v} \#r_{2}\\
      &\eta_{r_{2}} = \#r_{2} \xhra{ret\ R} \#\\
      &\eta_{s_{1}} = \#s_{1} \xhra{v:=42} \#s_{2}\\
      &\eta_{s_{2}} = \#s_{2} \xhra{ret\ S} \#
    \end{align*}
    
    \begin{minipage}[t]{2\linewidth}    
    Execution tree in extended notation\\
    \begin{tikzpicture}[node distance=1em]
      \node[]      (p1)     {$\spawn^{\eta_{p_{1}}}$};
      \node[right=of p1]      (p2)     {$\base^{\eta_{p_{2}}}$};
      \node[right=of p2]      (p3)     {$\use^{\eta_{p_{3}}}_{x,\bot}$};
      \node[below=of p1]      (q1)     {$\use^{\eta_{q_{1}}}_{x,\bot}$};
      \node[below=of p3]      (r1)     {$\base^{\eta_{r_{1}}}$};
      \node[right=of r1]      (r2)     {$\ret^{\eta_{r_{2}}}$};
      \node[below=of q1]      (s1)     {$\base^{\eta_{s_{1}}}$};
      \node[right=of s1]      (s2)     {$\ret^{\eta_{s_{2}}}$};      
      \node[right=1.4em of p3]      (np)     {$\nil^{\#}_{p_{4}}$};
      \node[right=1.4em of q1]      (nq)     {$\nil^{\#}_{q_{2}}$};
      \draw (p1) -- (p2) -- (p3) -- (r1) -- (r2) ;
      \draw (p1) -- (q1) -- (s1) -- (s2) ;
      \draw (q1) -- (nq);
      \draw (p3) -- (np);
    \end{tikzpicture}
    \end{minipage}
  \end{minipage}%
  \begin{minipage}[t]{.33\linewidth}  
    $\pi^{\tau} =
    \eta_{p_{1}}\eta_{p_{2}}\eta_{p_{3}}\eta_{r_{1}}\eta_{r_{2}}$\\
    $\pi^{1} = \eta_{q_{1}}\eta_{s_{1}}\eta_{s_{2}}$\\[1ex]

    Action tree\\
\begin{tikzpicture}[node distance=1em]
      \node[]      (p1)     {$\eta_{p_{1}}$};
      \node[right=of p1]      (p2)     {$\eta_{p_{2}}$};
      \node[right=of p2]      (p3)     {$\eta_{p_{3}}$};
      \node[below=of p1]      (q1)     {$\eta_{q_{1}}$};
      \node[right=of p3]      (r1)     {$\eta_{r_{1}}$};
      \node[right=of r1]      (r2)     {$\eta_{r_{2}}$};
      \node[right=of q1]      (s1)     {$\eta_{s_{1}}$};
      \node[right=of s1]      (s2)     {$\eta_{s_{2}}$};      
      \draw (p1) -- (p2) -- (p3) -- (r1) -- (r2) ;
      \draw (p1) -- (q1) -- (s1) -- (s2) ;
    \end{tikzpicture}\\[1ex]
    
    Execution tree\\
      \begin{tikzpicture}[node distance=1em]
      \node[]      (p1)     {$\eta_{p_{1}}$};
      \node[right=of p1]      (p2)     {$\eta_{p_{2}}$};
      \node[right=of p2]      (p3)     {$\eta_{p_{3}}$};
      \node[below=of p1]      (q1)     {$\eta_{q_{1}}$};
      \node[below=of p3]      (r1)     {$\eta_{r_{1}}$};
      \node[right=of r1]      (r2)     {$\eta_{r_{2}}$};
      \node[below=of q1]      (s1)     {$\eta_{s_{1}}$};
      \node[right=of s1]      (s2)     {$\eta_{s_{2}}$};      
      \node[right=1.4em of p3]      (np)     {$\#p_{4}$};
      \node[right=1.4em of q1]      (nq)     {$\#q_{2}$};
      \draw (p1) -- (p2) -- (p3) -- (r1) -- (r2) ;
      \draw (p1) -- (q1) -- (s1) -- (s2) ;
      \draw (q1) -- (nq);
      \draw (p3) -- (np);
    \end{tikzpicture}

\end{minipage}\\[1ex]
\rule{\linewidth}{1pt} 
  \caption{Construction of an Execution Tree}
  \label{fig:extreeConstr}
\end{figure} 

\autoref{fig:extreeConstr} depicts an example how a Java program can
be represented as a DPN and how, for a fixed execution, the execution
tree is build.  The flow graph model already models the synchronized
blocks as procedures $R$ and $S$.  The lock $x$ is bound to the call
of those procedures. We assume that the call \lstinline|q.start()|
spawns a thread that executes procedure $Q$. On the bottom left an
example execution which consists of the transition
$\eta_{p_{1}}\eta_{q_{1}}\eta_{s_{1}}\eta_{p_{2}}\eta_{s_{2}}\eta_{p_{3}}\eta_{r_{1}}\eta_{r_{2}}$
is shown.  On the right the construction of its execution tree via the
action tree is illustrated. On the bottom right the execution tree is
depicted in an extended notation which will be introduced a little
later.

We also overload $\post^{*}$ and $\lspost^{*}$ to
sets of execution trees:

\begin{definition}
  The lock-sensitive successors of a set $A$ of execution trees via
  $\Delta'\subseteq\Delta$ are
\[\lspost_{\Delta'}^{*}(A) =
\left\{\varcurlywedge (\pi\pi') \mid \pi,\pi\pi' \in \Pi(\mathcal M)
  ,\ \varcurlywedge (\pi) \in A,\ \delta(\pi') \subseteq \Delta'\right\}\]
The lock-insensitive version $\post_{\Delta'}^{*}$ is the one
corresponding to the lock-insensitive semantics.
\end{definition}

Note that different executions may produce the same execution tree,
because the tree does not capture the relative order of the
transitions of different threads completely.  In
\autoref{fig:extreeConstr} the execution tree also corresponds to the
four other lock-sensitive executions which reach the same
configuration. An execution that produces a given execution tree is
called a \emph{schedule} of that tree.  An execution tree which
possesses a schedule is called \emph{schedulable}.  If it possesses a
schedule in the lock-sensitive version of the semantics it is called
lock-sensitively schedulable. The configuration reached by an
execution can easily be reconstructed from the execution tree. We call
the corresponding function $\conf$ and overload it to sets of
execution trees.  This allows us to state the following adequacy
result for our definition of $\lspost^{*}$ for execution trees (the
analogous holds for $\post^{*}$):

\begin{proposition}
  For a reachable set of execution trees $A \subseteq
  \lspost^{*}_{\Delta}(\{p_{0}\gamma_{0}\})$ and
  $\Delta'\subseteq\Delta$ it holds that
  \begin{equation*}
    \lspost^{*}_{\Delta'}(\conf(A)) = \conf(\lspost^{*}_{\Delta'}(A))\ .
  \end{equation*}
\end{proposition}

We introduce additional notations for the nodes of an execution tree
to enhance readability and include some additional information.  As
noted earlier every node in an execution tree -- except the added
$p\gamma$ leafs -- corresponds to a transition of the DPN.  We denote
a node which corresponds to a transition $\eta$ of type \emph{base} as
$\base^{\eta}$.  A node corresponding to a \emph{call} transition
$\eta$ with one successor (i.e.\ a call which has no matching return
within the execution) is denoted as $\calln^{\eta}$. If it has two
successors (i.e.\ the call returns) it is denoted as
$\callr^{\eta}$. We denote by $\ret^{\eta}$ a node corresponding to a
\emph{return} transition $\eta$ and by $\nil_{\gamma}^{p}$ an inserted
$p\gamma$ leaf.  Nodes corresponding to \emph{monitor} transitions
$\eta$ are denoted analogously to calls as $\acq^{\eta}_{x,r}$ (the
lock is {\em finally acquired}, i.e., it remains acquired for the rest
of the execution) or $\use^{\eta}_{x,r}$ (the lock is {\em used},
i.e., it is released at some later point in the execution).  The
additional annotations $x$ and $r$ at monitor nodes denote the lock
$x$ and a bit $r \in \mathbb B = \{\bot,\top\}$ (interpreted as
\emph{false} and \emph{true}) indicating whether the operation is
reentrant, i.e., whether this lock is already acquired by the process
when executing the transition.  Nodes corresponding to \emph{spawn}
transitions $\eta$ are always binary and denoted as
$\spawn^{\eta}$. The left branch corresponds to the execution of the
newly spawned process and the right branch to the execution of the
spawning one.

\begin{figure}
  \centering
  \small
  \begin{minipage}{.8\linewidth}    
    \begin{tikzpicture}[node distance=1em]
      \node[]      (s)     {$\spawn^{\eta_{t_{2}.start()}}$};
      \node[below=4em of s]      (b1)     {$\acq^{\eta_{t_{2}:sync(b)}}_{b,\bot}$};
      
      \node[right=of b1]      (b2)     {$\base^{\eta_{else}}$};
      \node[right= of b2]      (b3)     {$\base^{\eta_{x=17}}$};
      \node[right =of b3]      (b5)     {$\nil^{p_{2}}_{\gamma_{t_{2}:print(x)}}$};
      
      \node[right =of s]      (a1)     {$\acq^{\eta_{t_{1}:sync(a)}}_{a,\bot}$};
      \node[right =of a1]      (a2)     {$\use^{\eta_{t_{1}:sync(b)}}_{b,\bot}$};
      \node[below=of a2]      (a3)     {$\base^{\eta_{x=42}}$};
      \node[right=of a3]      (a4)     {$\ret^{\eta_{rel(b)}}$};
      \node[right=of a2]      (a5)     {$\base^{\eta_{x=23}}$};
      \node[right=of a5]      (a6)     {$\nil^{p_{1}}_{\gamma_{t_{1}:write(x)}}$};
      \path[]  
      (s) edge (a1)
      (a1) edge (a2)
      (a2) edge (a3) edge (a5)
      (a3) edge (a4)
      (a5) edge (a6)
      (s) edge (b1)
      (b1) edge (b2)
      (b2) edge (b3)
      (b3) edge (b5)
      ;     
    \end{tikzpicture}
\end{minipage}
\vspace{1ex}

\begin{minipage}{.8\linewidth}    
    \begin{tikzpicture}[node distance=1em]
      \node[]      (s)     {$\spawn^{\eta_{t_{2}.start()}}$};
      \node[below=4em of s]      (b1)     {$\acq^{\eta_{t_{2}:sync(b)}}_{b,\bot}$};
      \node[right= of b1]      (b3)     {$\use^{\eta_{t_{2}:sync(a)}}_{a,\bot}$};
      \node[below=of b3]      (b4)     {$\base^{\eta_{\dots}}$};
      \node[right=of b4]      (b5)     {$\ret^{\eta_{t_{2}:rel(a)}}$};
      \node[right =of b3]      (b6)     {$\base^{\eta_{x=42}}$};
      \node[right =of b6]      (b7)     {$\nil^{p_{2}}_{\gamma_{t_{2}:write(x)}}$};
      
      \node[right =of s]      (a1)     {$\acq^{\eta_{t_{1}:sync(a)}}_{a,\bot}$};
      \node[right =of a1]      (a2)     {$\use^{\eta_{t_{1}:sync(b)}}_{b,\bot}$};
      \node[below=of a2]      (a3)     {$\base^{\eta_{\dots}}$};
      \node[right=of a3]      (a4)     {$\ret^{\eta_{rel(b)}}$};
      \node[right=of a2]      (a5)     {$\base^{\eta_{x=17}}$};
      \node[right=of a5]      (a6)     {$\nil^{p_{1}}_{\gamma_{t_{1}:print(x)}}$};
      \path[]  
      (s) edge (a1)
      (a1) edge (a2)
      (a2) edge (a3) edge (a5)
      (a3) edge (a4)
      (a5) edge (a6)
      (s) edge (b1)
      (b1) edge (b3)
      (b3) edge (b4)
      (b4) edge (b5)
      (b3) edge (b6)
      (b6) edge (b7)
      ;     
    \end{tikzpicture}
\end{minipage}
\caption[ ]{Two execution trees corresponding to the sixth and fifth
  example from \autoref{tab:ex1}.  }
  \label{fig:ex-tree1}
\end{figure}%

\autoref{fig:ex-tree1} contains two examples of execution trees which
correspond to an execution of some models of the sixth and fifth
example from \autoref{tab:ex1}. The left child of each binary node
grows downwards. The first execution tree is schedulable
lock-sensitively but the use of lock $b$ by the first thread must be
scheduled before the other thread acquires lock $b$.  The second
execution tree is reachable only in the lock-insensitive version of
the semantics, i.e., it is from the set
$\post_{\Delta}^{*}(\{\nil_{\gamma_{0}}^{p_{0}}\}) \setminus
\lspost_{\Delta}^{*}(\{\nil_{\gamma_{0}}^{p_{0}}\})$.  Here the main
thread wants to acquire $a$ and then use $b$.  The spawned thread
wants to acquire $b$ and then use $a$.  Obviously this is not possible
as one would unavoidably reach a deadlock.

Gawlitza {\it et al.}  \cite{join-lock-sens-fwd} showed that the set
of reachable execution trees for DPN without locks is effectively
tree-regular.  We recap their construction, slightly modified by
adding reentrance annotations, in \autoref{def:ta-dpn}.  Moreover,
they used acquisition structures to obtain a regular characterization
of reachable execution trees with (non-reentrant) well-nested locks
and extended this even further to \emph{joins} with spawned threads.
We pick up this approach and extend it to reentrant locks in
\autoref{prop:ls-reach}.  After introducing the concept of a
\emph{cut} of an execution tree, we then use acquisition and release
structures \cite{cav09-ps-dpn-trc} to also obtain an iterable
tree-regular characterization of reachable configurations. All this is
done by defining appropriate finite tree automata.

A finite (bottom-up) tree automaton $\mathcal{T}$ over a ranked
alphabet $\Sigma$ is a triple $(Q,Q_{f},\delta)$ consisting of finite
sets of states $Q$, accepting states $Q_{f}\subseteq Q$, and rules
$\delta$ of the form $f\ q_{1} \dots q_{n} \taarrow q$, where $f \in
\Sigma$ is a symbol of rank $n$ and $q,q_{1},\dotsc,q_{n} \in Q$.  If
a tree automaton can recognize the trees $t_{1},\dotsc,t_{n}$ with
states $q_{1},\dotsc,q_{n}$ respectively and it contains the above
rule, it can recognize the tree $f(t_{1},\dotsc,t_{n})$ with state
$q$.  A tree is accepted if it can be recognized with an accepting
state.  The language $\mathcal L(\mathcal T)$ is the set of all trees
accepted by the automaton.  A set of trees is called tree-regular if
it is the language of a finite tree automaton.  The intersection of
tree-regular sets is again tree-regular (accepted by the product
automaton) and emptiness of the language of a given tree automaton is
decidable in linear time.  For an introduction to tree automata we
refer the reader to Comon {\it et al.} \cite{tata2007}.  When defining
concrete automata we leave out sub- and superscripts of nodes which do
not influence the automata and write an underscore to denote arbitrary
values, e.g., $\nil \taarrow \_$ denotes $\nil^{p}_{\gamma} \taarrow
U$ for all $p$, $\gamma$ and $U$.

\begin{definition} 
  \label{def:ta-dpn}
  For a Monitor-DPN $\mathcal{M}$ we define the tree automaton
  $\mathcal{T}_{\mathcal{M}}$, which is intended to accept the \emph{
    lock-insensitive execution trees} of $\mathcal{M}$ .  The state
  space is $P \times \Gamma \times P \times \{\bot,\top\} \times
  2^{X}$ and the accepting states are $\{p_0\}\times\{\gamma_0\}\times
  P \times \{\bot,\top\} \times\{\emptyset\}$.  For all $\eta \in
  \Delta$, $t\in \{\bot,\top\}$, $p,p',p''.p''' \in P$, $\gamma,\gamma',\gamma''\in \Gamma$ and $ls
  \subseteq X$ the automaton contains the following rules.  We write
  $\frac{f\ q_{1} \dots q_{n}}{q}$ to denote the rule $f\ q_{1} \dots
  q_{n} \taarrow q$.

  \noindent
   \begin{align*} 
    \small
      \tfrac{\nil_\gamma^{p} }{ (p,\gamma,p,\bot,ls)}&, \\
      \tfrac{\ret^{\eta} }{ (p,\gamma,p',\top,ls) } &\quad\text{ if } \eta = p\gamma \xhra{a} p',  \\
      \tfrac{\base^{\eta}\ (p',\gamma',p'',t,ls) }{ (p,\gamma,p'',t,ls) } &\quad\text{ if } \eta = p\gamma \xhra{a} p'\gamma',  \\
      \tfrac{\calln^{\eta}\ (p',\gamma',p'',\bot,ls) }{ (p,\gamma,p'',\bot,ls) } &\quad\text{ if } \eta = p\gamma \xhra{a} p'\gamma'\gamma'',  \\
      \tfrac{\callr^{\eta}\ (p',\gamma',p'',\top,ls)\
      (p'',\gamma'',p''',t,ls)}{   (p,\gamma,p''',t,ls) } &\quad\text{ if } \eta =  p\gamma \xhra{a} p'\gamma'\gamma'', \\
      \tfrac{\spawn^{\eta}\  (p',\gamma',\_,\_,\emptyset)\ (p'',\gamma'',p''',t,ls)}{ (p,\gamma,p''',t,ls) } &\quad\text{ if } \eta =  p\gamma \xhra{a} p'\gamma'p''\gamma'',\\[0.5em]
      \tfrac{\acq^{\eta}_{x,r}\ (p',\gamma',p'',\bot,ls \cup \{x\})
      }{
      (p,\gamma,p'',\bot,ls) } &\quad\text{ if } \eta =  p\gamma \xhra{a,x} p'\gamma'\gamma'',\ r = \top
      \Leftrightarrow x \in ls, \\
      \tfrac{\use^{\eta}_{x,r}\ (p',\gamma',p'',\top,ls \cup \{x\})\
      (p'',\gamma'',p''',t,ls)}{ (p,\gamma,p''',t,ls) } &\quad\text{ if
      } \eta = p\gamma \xhra{a,x} p'\gamma'\gamma'', \ r = \top
      \Leftrightarrow x \in ls.
    \end{align*}
\end{definition}

For a tree recognized by this automaton the corresponding state
records in the first two places the control state and top of stack
with which the execution started.  That is the head of the rule
corresponding to the root node.  The third place records the control
state that has been reached by the main thread with its last action,
that is the control state annotated at the right-most $\nil$ or $\ret$
node.  Whether this last action of the main thread is a $\ret$ node or
not is marked in the forth place by $\top$ or $\bot$ respectively.
Moreover, it contains a set of locks in the fifth place, which is used
to check the marking of reentrant lock operations.  That is, it
calculates for each node which locks the thread holds upon the
execution of the corresponding transition.  In the lock-insensitive
setting this tree automaton already characterizes the set of reachable
execution trees:

\begin{proposition}
  \label{lem:reach}
 $\mathcal{L}(\mathcal{T_{M}})
  = \post^{*}_{\Delta}(\{\nil_{\gamma_{0}}^{p_{0}}\})$\ .
\end{proposition}

To obtain $\post^{*}_{\Delta'}$ for arbitrary $\Delta'\subseteq\Delta$
one can utilize the following simple tree automaton with a single state
to check that only transition from $\Delta'$ are used:

\begin{definition}
  The tree automaton $\mathcal{T}_{\Delta'}$ has the single state
  $\bullet$ (which is accepting) and the following rules for all $\eta
  \in \Delta'$:
  \begin{align*}
    \nil  &\taarrow \bullet&
    \ret^{\eta} &\taarrow \bullet&
\calln^{\eta}\ \bullet &\taarrow \bullet&
\acq^{\eta}\ \bullet &\taarrow \bullet\\
\callr^{\eta}\ \bullet\ \bullet &\taarrow \bullet&
\use^{\eta}\ \bullet\ \bullet &\taarrow \bullet&
\spawn^{\eta}\ \bullet\ \bullet &\taarrow \bullet
  \end{align*}
\end{definition}

\begin{proposition}
  \label{lem:reach-Delta}
 $\mathcal{L}(\mathcal{T_{M}}) \cap \mathcal{L}(\mathcal{T}_{\Delta'})
  = \post^{*}_{\Delta'}(\{\nil_{\gamma_{0}}^{p_{0}}\})$\ .
\end{proposition}

\subsection{Lock-Sensitive Analysis}
\label{sec:lock-sens-analys}
As $\lspost^{*}_{\Delta'}(\{\nil_{\gamma_{0}}^{p_{0}}\}) \subseteq
\post^{*}_{\Delta'}(\{\nil_{\gamma_{0}}^{p_{0}}\})$ one can obtain the
set of lock-sensitively reachable execution trees by filtering out
those which do not have a lock-sensitive schedule.  This is done by
the acquisition structure, which precisely models the restrictions to
a possible schedule imposed by the final acquisitions and uses of
locks. Only non-reentrant actions need to be considered: Reentrant
actions can always be executed, and their execution cannot inhibit
execution of other threads.

\begin{definition}
  \label{def:ta-ah}
  The tree automaton $\mathcal{T}_{ah}$ accepts from
  the set of lock-insensitively reachable execution trees those which
  possess a lock-sensitive schedule.  The state space is $2^{X} \times
  2^{X} \times 2^{X \times X}$ with accepting states $2^{X} \times
  2^{X} \times \{G \in 2^{X \times X} \mid G \text{ is acyclic}\}$.
  We write sets of nodes to denote a rule for each node of the given
  set.  The rules are as follows for all $x\in X$:
  
  \noindent
  \footnotesize  
  \begin{gather*}    
    \nil \taarrow (\emptyset,\emptyset,\emptyset) \quad
    \ret \taarrow (\emptyset,\emptyset,\emptyset) \\
    \base\ \alpha \taarrow  \alpha \quad 
    \calln\ \alpha \taarrow \alpha \quad
    \acq_{x,\top}\ \alpha  \taarrow \alpha \\
    \frac{ \{\callr,\use_{x,\top},\spawn\}\ (\ASet,\USet,\GSet)\ (\ASet',\USet',\GSet') }{
      (\ASet \cup \ASet',\ \USet \cup \USet',\ \GSet \cup \GSet') } \text{ if } 
    \ASet \cap \ASet' = \emptyset\\
    \frac{\use_{x,\bot}\ (\ASet,\USet,\GSet)\ (\ASet',\USet',\GSet') }{
      (\ASet \cup \ASet',\ \USet \cup \USet' \cup \{x\},\ \GSet \cup
      \GSet') } \text{ if } 
    \ASet \cap \ASet' = \emptyset\\
    \frac{\acq_{x,\bot}\ (\ASet,\USet,\GSet) }{
      (\ASet \cup \{x\},\ \USet \cup \{x\},\ \GSet \cup \{(x,u)|u \in \USet \}) }
     \text{ if }  x \notin \ASet
\end{gather*}
\end{definition}

The first component of the state of this tree automaton represents the
set of locks which are finally acquired non-reentrantly within the
subtree, that means, there is a corresponding $\acq$-node.  It is used
to assert that (1) no lock is finally acquired twice, which is
achieved by the side conditions of the above rules.  The second
component represents all locks which are used (non-reentrantly,
including final acquisitions) in the subtree.  The third component is
the {\em acquisition graph}, which represents order constraints
imposed by the program order.  It contains an edge $x \rightarrow y$
if in any schedule the lock $y$ must be used or acquired
non-reentrantly after the lock $x$ has been finally acquired
non-reentrantly.  The acceptance condition ensures that (2) the
acquisition graph is acyclic.  Criteria (1) and (2) suffice to
precisely characterize the set of lock-sensitively schedulable
execution trees \cite{cav09-ps-dpn-trc}:
\begin{proposition} 
\label{prop:ls-reach}
$\mathcal{L}(\mathcal{T_{M}}) \cap \mathcal{L}(\mathcal{T}_{ah}) =
\lspost^{*}_{\Delta}(\{\nil_{\gamma_{0}}^{p_{0}}\})$\ .
\end{proposition}

The corresponding version restricted to a set $\Delta'\subseteq\Delta$
of allowed transition is:

\begin{proposition} 
\label{prop:ls-reach-delta}
$\mathcal{L}(\mathcal{T_{M}}) \cap \mathcal{L}(\mathcal{T}_{ah}) \cap
\mathcal{L}(\mathcal{T}_{\Delta'}) =
\lspost^{*}_{\Delta'}(\{\nil_{\gamma_{0}}^{p_{0}}\})$\ .
\end{proposition}

One easily sees that if a tree is not accepted by the tree automaton
$\mathcal T_{ah}$ it cannot be scheduled: Either some lock is finally
acquired twice by different threads or there exists a cycle in the
acquisition graph corresponding to an unavoidable deadlock.  On the
other hand, if a tree is accepted one can show that it has a schedule.
In this schedule the final acquisitions are executed as late as
possible in a topological ordering of the acyclic acquisition graph
and the uses of locks are scheduled atomically between the final
acquisitions. The latter is possible due to the requirements captured
in the acquisition graph and as a thread holds the same locks before
and after a use due to well-nestedness \cite{join-lock-sens-fwd}.

The language $\mathcal{L}(\mathcal{T_{M}}) \cap
\mathcal{L}(\mathcal{T}_{ah})$ is the language of the product
automaton of $\mathcal{T_{M}}$ and $\mathcal{T}_{ah}$.  This can be
used to answer reachability questions for tree-regular sets of
configurations by checking their intersection for emptiness.  

One may ask why we used reentrance annotations instead of calculating
the required information directly in the last automaton.  Note that
the first tree automaton ($\mathcal{T_{M}}$) is deterministic when
evaluated top down and the second one ($\mathcal{T}_{ah}$) when
evaluated bottom up, but both are extremely nondeterministic when
evaluated in the opposite direction.  We exploit this with a
specialized emptiness check which evaluates one automaton only bottom
up and the other top down.  We will report on this in
Section~\ref{sec:dpn-analysis}.  This determinism of
$\mathcal{T}_{ah}$ would be lost if it would calculate reentrance
information.
\begin{example}
  \label{ex:data-race}
  A simple application is calculation of may-happen-in-parallel
  information.  That is, given two sets $R$ and $W$ of stack symbols,
  is it possible that two threads simultaneously reach a topmost stack
  symbol of the corresponding sets?  The relevant execution trees are
  accepted by the tree automaton $\mathcal{T}_{rw}$ with state space
  $2^{\{r,w\}}$, single accepting state $\{r,w\}$ and the following
  rules:

  \noindent
  {\footnotesize
  \begin{gather*}
    \begin{aligned}
      \nil_{\gamma} &\taarrow \{r\} \text{ for } \gamma \in R\quad&
      \nil_{\gamma} &\taarrow \{w\}  \text{ for } \gamma \in W\quad&
      \nil_{\gamma} &\taarrow \emptyset \text{ for } \gamma \notin (R \cup W)
    \end{aligned}\\
      \ret \taarrow \emptyset \quad \{\acq,\base, \calln\}\ \alpha \taarrow \alpha\\
    \{ \callr, \spawn, \use \}\ \alpha\ \beta \taarrow \alpha \cup \beta
  \end{gather*}}

By checking whether $\mathcal{L}(\mathcal{T_{M}}) \cap
\mathcal{L}(\mathcal{T}_{ah}) \cap \mathcal{L}(\mathcal{T}_{rw})
\overset{?}{=} \emptyset$ one can 
prove the absence of a data race on some variable if $R$ represents
the set of stack symbols where the variable is read or written and $W$
the set of stack symbols where the variable is written.  This rules
out all the spurious data races in the examples in \autoref{tab:ex1},
i.e., all except the two actual races between the assignment
\lstinline|x = 23| by the first and the accesses to \lstinline|x| by
the other thread in the sixth example.  A schedulable execution tree
witness is depicted in the upper part of \autoref{fig:ex-tree1}.
\end{example}

\subsection{Iterable Analysis}
\label{sec:iterable-analysis}
The hitherto presented techniques allow only to check for reachability
from the initial configuration of the Monitor-DPN.  We now extend this
technique to answer reachability queries starting from arbitrary
tree-regular sets of reachable execution trees.  We obtain a
tree-regular representation of $\lspost_{\Delta'}^{*}(A)$ for
arbitrary tree-regular $A \subseteq
\lspost_{\Delta}^{*}(\{\nil_{\gamma_{0}}^{p_{0}}\})$ and $\Delta'
\subseteq \Delta$.  This in particular allows us to answer iterated
reachability questions: Given a sequence $A_{1},\dotsc, A_{n}$ of
regular sets of execution trees and sets of DPN transitions
$\Delta_{1},\dotsc,\Delta_{n} \subseteq \Delta$, we compute $A_{n}
\cap \lspost_{\Delta_{n}}^{*}(A_{n-1}\cap
\lspost_{\Delta_{n-1}}^{*}(\dots A_{1}\cap
\lspost_{\Delta_{1}}^{*}(\{\nil_{\gamma_{0}}^{p_{0}}\}) \dots ))$ and
check it for emptiness, in order to decide whether there is an
execution visiting the set $A_{1}, \dotsc, A_{n}$ in the given order
using the given transitions.

In order to compute $\lspost^{*}_{\Delta'}(A)$, we exploit the fact
that all execution trees of prefixes of an execution are closely
related to prefixes of the final tree: They are obtained by cutting
the final tree, replacing subtrees by $\nil$-nodes and changing some
returning calls or uses to non-returning calls or acquisitions as
necessary. Note however, that in the presence of locking not all
prefixes of an execution tree correspond to prefixes of its schedules.
We use an additional type of node (a $\cut$-node) to \emph{mark} an
intermediate configuration in the tree. For each thread that existed
at the intermediate configuration, a cut node marks the position after
its last step before reaching the intermediate configuration.

In a first step we characterize the execution-trees which are
wellformed with respect to the placement of cut nodes.  In order to
avoid several additional case distinctions we make the assumption that
no thread may empty its stack, that is the corresponding procedure
with which its execution started may not perform a return action.
With this assumption, in the lock-insensitive setting, it suffices to
check that each thread spawned \emph{before} the cut contains exactly
one cut node and that no thread spawned afterwards contains any cuts.
This can easily be checked by a tree automaton.  To obtain the regular
representation we first extend $\mathcal{T_{M}}$ by the
rule $\cut_{\gamma}^{p}\ (p, \gamma, p', t,ls) \taarrow (p, \gamma,
p', t,ls)$.  This allows arbitrary (well-annotated) cut-nodes to be
inserted in the execution tree.  The following tree automaton checks
that the inserted cut nodes mark a legit intermediate configuration.

\begin{definition}
  \label{def:ta-cwf}
  The cut wellformed trees are accepted by the tree automaton
  $\mathcal{T}_{cwf}$ which has the state space
  $\{\bot,\top,\sptop,\spbot\}$, accepting state $\top$ and the
  following rules.  We write sets of states to denote a corresponding
  rule for every state of the given set.  We leave out the rules for
  $\use$/$\acq$ nodes as these are the same as for call nodes.

  \noindent
  {\footnotesize
    \begin{align*}
      \callr\ \{\bot,\sptop\}\ \top &\taarrow \top 
      &\callr\ \top\ \{\bot,\spbot\} &\taarrow \top
      &\nil &\taarrow \bot \\
      \callr\ \sptop\ \{\bot,\sptop\}  &\taarrow \sptop 
      & \callr\ \spbot\ \{\bot,\spbot\} &\taarrow \spbot
      &\ret &\taarrow \bot\\
      \callr\ \bot\ \sptop &\taarrow \sptop 
      & \callr\ \bot\ \spbot &\taarrow \spbot
      &\base\ q & \taarrow q\\
      \callr\ \bot\ \bot &\taarrow \bot
      & \spawn\ \top\ \{\bot,\sptop\}  &\taarrow \sptop 
      & \calln\ q & \taarrow q\\
      \spawn\ \{\bot,\spbot\}\ \{\bot,\spbot\}  &\taarrow \spbot   
      &\spawn\ \top\ \top  &\taarrow \top 
      &\cut\ \{\bot,\spbot\} &\taarrow \top 
    \end{align*}}
\end{definition}

The intuition behind the rules is as follows: A tree recognized with
the accepting state $\top$ is cut-wellformed.  A tree recognized with
$\bot$ does not contain any cut or spawn nodes.  A tree recognized
with state $\sptop$ contains some spawn nodes and the trees
corresponding to the spawned threads are cut-wellformed, but the main
thread of the tree is missing its cut node.  In this case the tree
must be in the left (returning) branch of a returning call/use and the
corresponding cut node must be in the right branch.  Finally $\spbot$
corresponds to a tree which contains spawn nodes but no cut nodes.
This requires that it is located \emph{after} the cut.  By careful
inspection of the individual cases one can see that the above tree
automaton calculates the correct information.

Based on a cut-wellformed execution tree we need to check whether the
execution tree corresponding to the execution marked by the cut is
from the given tree-regular set for which we want to calculate
$\post^{*}$.  We can do this by using a tree transducer to obtain the
execution tree in question.  A tree transducer can be seen as a tree
automaton with output.  The rules have the form $f\ q_{1}(t_{1})\dots
q_{n}(t_{n}) \taarrow q(f'\ t_{i_{1}}\dots t_{i_{k}})$. The semantics
is similar to the one for tree automata except that the states are
augmented with an output which is inductively constructed from the
output of the subtrees in each rule.  It is known that the inverse
image of a tree-regular set under a tree transducer is again
tree-regular (Engelfriet \cite{springerlink:10.1007/BF01704020}).

\begin{definition}
  \label{def:ta-ct}
  The tree transducer $\mathcal{T}_{ct}$ with state space
  $\{\bot,\top\}$, accepting state $\top$ and the following rules
  obtains the execution tree marked by the cut from a well-cut
  execution tree.

  \noindent
  {\footnotesize
    \begin{align*}
      \nil_{\gamma}^{p} &\taarrow \bot(\nil_{\gamma}^{p}) &   \ret^{\eta} &\taarrow \bot(\ret^{\eta})\\
      \base^{\eta}\ q(w) & \taarrow q(\base^{\eta}\ w) &
      \calln^{\eta}\ q(w) & \taarrow q(\calln^{\eta}\ w)\\
      \callr^{\eta}\ \top(c)\ \bot(\_) &\taarrow \top(\calln^{\eta}\ c)
      &\callr^{\eta}\ \bot(c)\ \bot(t) &\taarrow \bot(\callr^{\eta}\ c\ t))\\
      \acq^{\eta}_{x,r}\ q(w) & \taarrow q(\acq^{\eta}_{x,r}\ w)&
      \use^{\eta}_{x,r}\ \top(c)\ \bot(\_) &\taarrow \top(\acq^{\eta}_{x,r}\ c)\\
      \use^{\eta}_{x,r}\ \bot(c)\ \bot(t) &\taarrow \bot(\use^{\eta}_{x,r}\ c\ t))&
      \spawn^{\eta}\ \_(c)\ q(t) &\taarrow q(\spawn^{\eta}\ c\ t) \\
      \cut_{\gamma}^{p}\ \bot(\_) &\taarrow \top(\nil_{\gamma}^{p})
    \end{align*}
  } 
\end{definition}

\begin{proposition}
  \label{lem:cwf2}
  For tree-regular $A \subseteq
  \post^{*}_{\Delta}(\{\nil_{\gamma_{0}}^{p_{0}}\})$ it holds that
  \begin{equation*}
    \post^{*}_{\Delta}(A) = (\mathcal{L}(\mathcal{T_{M}}) \cap
    \mathcal{L}(\mathcal{T}_{cwf}) \cap
    \mathcal{T}_{ct}^{-1}(A))|_{\cut} \ .
  \end{equation*}
\end{proposition}

Here, $|_{\cut}$ denotes the projection which removes all cut nodes
from the tree which can be done by a simple tree transducer.  

In order to compute $\post_{\Delta'}^{*}(A)$ for arbitrary $\Delta'
\subseteq \Delta$ we again use a simple tree automaton on the set
of execution trees with cut nodes, which checks that only rules from
$\Delta'$ are used \emph{after} the cut.

\begin{definition}
  For a set $\Delta' \subseteq \Delta$ the tree automaton
  $\mathcal{T}^{c}_{\Delta'}$ checks that only rules from $\Delta'$ are
  used \emph{after} the cut.  The state space is $\{\bot,+,\top \}$
  and $\top$ is the only accepting state.  For the definition of the
  rules let $\eta$ range over $\Delta$, $\eta_{g}$ range over
  $\Delta'$, and $\eta_{b}$ range over $\Delta\setminus \Delta'$.  The
  rules are as follows:
{\small
  \begin{align*}    
    \base^{\eta_{g}}\ q & \taarrow q &
    \base^{\eta_{b}}\ \{\bot,+\}& \taarrow +&
    \base^{\eta_{b}}\ \top & \taarrow \top\\
    \callr^{\eta}\ \top\ \bot& \taarrow \top&
    \callr^{\eta}\ \{\bot,+\}\ \top& \taarrow \top\\
    \{\spawn^{\eta_{b}},\callr^{\eta_{b}}\}\ \bot\ \bot& \taarrow +&
    \{\spawn^{\eta_{g}},\callr^{\eta_{g}}\}\ \bot\ \bot& \taarrow \bot\\
    \{\spawn^{\eta},\callr^{\eta}\}\ +\ \{\bot,+\}& \taarrow +&
    \{\spawn^{\eta},\callr^{\eta}\}\ \bot\ +& \taarrow +\\
    \spawn^{\eta_{g}}\ \top\ \bot& \taarrow \bot&
    \spawn^{\eta_{b}}\ \top\ \bot& \taarrow +&
    \spawn^{\eta}\ \top\ \top& \taarrow \top\\
    \spawn^{\eta}\ \top\ +& \taarrow +&
    \cut\ \bot & \taarrow \top
  \end{align*}
}
  The rules for $\acq$ and $\calln$-nodes are the same as for
  $\base$-nodes and the rules for $\use$-nodes are the same as those
  for $\callr$-nodes.
\end{definition}

Here the state $\bot$ indicates that there is neither a cut node in
the main thread nor a transition that is not from $\Delta'$ in any
thread.  The state $+$ indicates that there is a transition not from
$\Delta'$ which may lay below the cut as there is no cut node in the
main thread.  Therefore at a cut node only $\bot$ is accepted and the
state changes to $\top$ which indicates that there is a cut in the
main thread and afterwards only actions from $\Delta'$ are used.  As
the left branch of a returning call is executed before the right
branch, if there is a cut node in the left branch of such a call, then
all transitions in the right branch must be from $\Delta'$. Thus, only
$\bot$ is accepted for the right branch.

\begin{proposition}
  \label{lem:cwf2delta}
  For tree-regular $A \subseteq
  \post^{*}_{\Delta}(\{\nil_{\gamma_{0}}^{p_{0}}\})$ and $\Delta'
  \subseteq \Delta$ it holds that
  \begin{equation*}
    \post^{*}_{\Delta'}(A) = (\mathcal{L}(\mathcal{T_{M}}) \cap
    \mathcal{L}(\mathcal{T}_{cwf}) \cap \mathcal{T}_{ct}^{-1}(A) \cap
    \mathcal{L}(\mathcal{T}^{c}_{\Delta'}))|_{\cut} \ .
  \end{equation*}
\end{proposition}

\subsection{Iterable Lock-Sensitive Analysis}
\label{sec:iter-lock-sens}
\begin{figure}
    \footnotesize
  \centering
  \pgfdeclarelayer{bg}    
  \pgfsetlayers{bg,main}  
  \begin{minipage}{16em}    
      \begin{tikzpicture}[node distance=1em]
        \node[]      (s1)     {$\spawn$};
      \node[right= of s1,fill=white]      (c1)     {$\cut$};
      \node[right= of c1]      (u1)     {$\use_x$};
      \node[below=.6em of u1]      (t1)     {$\ret$};
      \node[right=of u1]      (n1)     {$\nil$};
      \node[below=3em of s1]      (a2)     {$\acq_x$};
      \node[right=of a2,fill=white]      (c2)     {$\cut$};
      \node[right=of c2]      (n2)     {$\nil$};
        
      \path[]  
      (s1)    edge[]        node[]       {} (c1) 
      (c1)    edge[]        node[]       {} (u1) 
      (u1)    edge[]        node[]       {} (t1) 
      (u1)    edge[]        node[]       {} (n1) 

      (s1)    edge[]        node[]       {} (a2)
      (a2)    edge[]        node[]       {} (c2) 
      (c2)    edge[]        node[]       {} (n2)
      ; 
      \begin{pgfonlayer}{bg}
        \draw[shorten >=-2.5em,shorten <=-2.5em, dashed,blue] (c1) --
        (c2);
      \end{pgfonlayer}
    \end{tikzpicture}%
  \end{minipage}\rule{2em}{0em}%
  \begin{minipage}{20em}    
    \begin{tikzpicture}[node distance=1em]        
      \node[]      (s1)     {$\spawn$};
      \node[right=5em of s1]      (u1)     {$\use_y$};
      \node[right=of u1]      (n1)     {$\nil$};
      \node[below=of u1,fill=white]      (c1)     {$\cut$};
      \node[right=of c1]      (u2)     {$\use_x$};
      \node[right=of u2]      (r1)     {$\ret$};
      \node[below=.6em of u2]      (r2)     {$\ret$};
      
      \node[below=of s1]      (u3)     {$\use_x$};
      \node[right=of u3]      (n2)     {$\nil$};
      \node[below=of u3,fill=white]      (c2)     {$\cut$};
      \node[right=of c2]      (u4)     {$\use_y$};
      \node[right=of u4]      (r3)     {$\ret$};
      \node[below=.6em of u4]      (r4)     {$\ret$};
        
      \path[]  
      (s1)    edge[]        node[]        {} (u1) 
              edge[]        node[]       {} (u3) 
      (u1)    edge[]        node[]        {} (n1) 
              edge[]        node[]       {} (c1) 
      (c1)    edge[]        node[]        {} (u2) 
      (u2)    edge[]        node[]        {} (r1) 
              edge[]        node[]        {} (r2) 
      (u3)    edge[]        node[]        {} (n2) 
              edge[]        node[]       {} (c2) 
      (c2)    edge[]        node[]        {} (u4) 
      (u4)    edge[]        node[]        {} (r3) 
          edge[]        node[]       {} (r4) 
      ;     

        \begin{pgfonlayer}{bg}
        \draw [dashed,blue] plot [smooth] coordinates {
          ([xshift=-1ex,yshift=-1.5ex]c2.south west) 
          ([yshift=1ex]n2.west) 
          (n2.north east)
          ([yshift=-1ex]c1.west)
          (c1.north east) 
          ([xshift=-1ex,yshift=2ex]n1.west)};
      \end{pgfonlayer}
    \end{tikzpicture}
  \end{minipage}
  
  \caption[ ]{Not lock-sensitively schedulable execution trees with cut.}
   \label{fig:ex-tree-cut}
\end{figure}
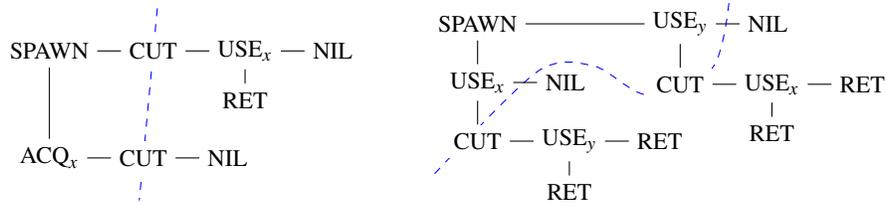

In the lock-sensitive case, it remains to be checked whether the final
configuration can be reached lock-sensitively from the intermediate
configuration marked by the cut.  This clearly requires that the final
configuration is lock-sensitively reachable.  Note that we assume that
the intermediate configuration is already lock-sensitively reachable.
Yet this is not sufficient, as reaching the cut may introduce
additional constraints.  Two prototypical examples are shown in
\autoref{fig:ex-tree-cut}, where the dashed lines illustrate how cut
nodes divide an execution tree into two parts.  In both trees the
intermediate configuration marked by the cut and the final
configuration are lock-sensitively reachable, but the final
configuration is not lock-sensitively reachable from the intermediate
one.  To obtain a sufficient criterion we additionally use {\em
  release structures}, which, analogously to acquisition structures
for acquisitions, model restrictions for releasing held locks
\cite{cav09-ps-dpn-trc}.  Firstly, one must ensure that all locks used
at the cut can actually be released afterwards.  This is not always
possible as releasing one lock may require using another one first as
depicted in the right part of \autoref{fig:ex-tree-cut}. Secondly, one
must check that no lock finally acquired before the cut is used by
another thread after the cut, as depicted in the left part of
\autoref{fig:ex-tree-cut}.  A difference to the earlier approach
\cite{cav09-ps-dpn-trc} is that there the acquisition graph is
computed independently for each phase, while we check the acquisition
graph of the whole execution tree.  It is easy to see that this
approach is equivalent under the assumption that the previous phases
are reachable lock sensitively and that no lock finally acquired
before the cut is used afterwards.  The reason for this approach is
that it is easier for a tree automaton to calculate this information
because it does not need to make additional distinctions whether it is
above or below the cut.  Moreover we can reuse the above presented
tree automaton $\mathcal{T}_{ah}$ by simply ignoring the cut nodes,
that is we add the rule $\cut q \taarrow q$.  The additional
constraints can be checked with the automaton given in
Definition~\ref{def:ta-relstr}~and~\ref{def:ta-acq-use}.

\begin{definition}
  \label{def:ta-relstr}
  The release structure is computed by the following tree automaton
  $\mathcal{T}_{rh}$ with state space $\mathbb{B}\times 2^{X} \times
  2^{X\times X}$, accepting states $\mathbb{B}\times 2^{X} \times \{G
  \in 2^{X\times X} \mid G \text{ is acyclic}\}$, and transitions:
 
  \noindent
  \footnotesize
  \begin{gather*}
    \begin{aligned}
      \nil &\taarrow (\bot,\emptyset,\emptyset) &
      \ret &\taarrow (\bot,\emptyset,\emptyset)\\
      \base\ p &\taarrow p&
      \calln\ p &\taarrow p\\
      \acq_{x,\_}\ p &\taarrow p& \cut\ (\bot,U,G) &\taarrow
      (\top,U,G)
    \end{aligned}\\[.5ex]
    \begin{aligned}
      \callr\ (C,U_c,G_c)\ (\bot,U_r,G_r) &\taarrow (C,U_c \cup U_r,G_c \cup G_r)\\
      \callr\ (\bot,\_,G_c)\ (\top,U_r,G_r) &\taarrow (\top,U_r,G_c
      \cup G_r)
    \end{aligned}\\[.5ex]
    \begin{aligned}
      \use_{x,\top}\ (C,U_c,G_c)\ (\bot,U_r,G_r) &\taarrow (C,U_c \cup U_r,G_c \cup G_r)\\
      \use_{x,\_}\ (\bot,\_,G_c)\ (\top,U_r,G_r) &\taarrow (\top,U_r,G_c \cup G_r)\\
      \use_{x,\bot}\ (\bot,U_c,G_c)\ (\bot,U_r,G_r) &\taarrow (\bot,U_c \cup U_r \cup \{x\},G_c \cup G_r)\\
      \use_{x,\bot}\ (\top,U_c,G_c)\ (\bot,U_r,G_r) &\taarrow
      (\top,U_c \cup U_r,G_c \cup G_r \cup \{(y,x) \mid y \in U_c \})
    \end{aligned}\\[.5ex]
    \begin{aligned}
      \spawn\ (\_,\_,G_s)\ (C,U_r,G_r) &\taarrow (C,U_r,G_s \cup
      G_r)
    \end{aligned}
  \end{gather*} 
\end{definition}

In this automaton the bit in the first place is set to top once the
main thread of the tree contains a cut node.  The lock-set in the
second place contains the locks which are used by the main thread of
the subtree after the cut.  The lock-set is discarded if it becomes
apparent that the subtree lies over the cut.  The release-graph in the
third place contains an edge from $y$ to $x$ if there is a use of
$x$ inside which there is the cut and, after the cut, a use of $y$.
This means that $y$ needs to be available before $x$ can be released.

\begin{definition}
  \label{def:ta-acq-use} The tree automaton $\mathcal{T}_{ht}$ checks
  that no lock finally acquired before the cut is used afterwards.  It
  has the state space $\mathbb{B} \times 2^{X} \times 2^{X}$. The
  accepting states are $\{(C,A,U) \in \mathbb{B} \times 2^{X}
  \times 2^{X} \mid A \cap U = \emptyset\}$. The rules are the
  following. Reentrant use nodes are handled in the same way as
  returning calls.
  
  \noindent
  { \footnotesize 
    \begin{minipage}{0.35\linewidth}
      \begin{align*}
        &\nil \taarrow (\bot,\emptyset,\emptyset)\\
        &\ret \taarrow (\bot,\emptyset,\emptyset)\\
        &\base\ p \taarrow  p\\
        &\calln\ p \taarrow p\\
        &\acq_{x,\top}\ p  \taarrow p\\
        &\cut\ (\bot,A,U) \taarrow (\top,A,U)
      \end{align*}
    \end{minipage}%
    \hspace{-0.07\linewidth}%
    \begin{minipage}{0.7\linewidth}
      \begin{align*}
        \callr\ (\bot,A,U)\ (\bot,A',U') &\taarrow
        (\bot, \ A \cup A',\ U \cup U')\\
        \callr\ (\top,A,U)\ (\bot,\_,U') &\taarrow
        (\top,\ A ,\ U \cup U')\\
        \callr\ (\bot,A,\_)\ (\top,A',U') &\taarrow
        (\top,\ A \cup A',\ U')\\
        \use_{x,\bot}\ (\bot,A,U)\ (\bot,A',U') &\taarrow
        (\bot,\ A \cup A',\ U \cup U' \cup \{x\})\\
        \use_{x,\bot}\ (\top,A,U)\ (\bot,\_,U') &\taarrow
        (\top,\ A ,\ U \cup U')\\
        \use_{x,\bot}\ (\bot,A,\_)\ (\top,A',U') &\taarrow
        (\top,\ A \cup A',\ U')\\
        \acq_{x,\bot}\ (\bot,A,U) &\taarrow
        (\bot,\ A,\ U \cup \{x\})\\
        \acq_{x,\bot}\ (\top,A,U) &\taarrow
        (\top,\ A \cup \{x\},\ U)\\
        \spawn\ (\_,A,U)\ (C,A',U') &\taarrow (C,\ A \cup A',\ U \cup
        U')
      \end{align*}    
    \end{minipage}
  }
\end{definition}

Again, the bit in the first position of a state indicates whether the
main thread has seen a cut.  The first lock-set represents the locks
that are finally acquired before the cut.  The second lock-set
represents the locks used or finally acquired after the cut similar to
those in the previous automaton, except that we include uses from
spawned threads.

Using these automata we can characterize the set of execution trees
that are lock-sensitively reachable from a regular set of lock-sensitively
reachable execution trees:

\begin{proposition}
  \label{lem:it-reach} For tree-regular $A \subseteq
  \lspost_{\Delta}^{*}(\{\nil_{\gamma_{0}}^{p_{0}}\})$ and $\Delta'
  \subseteq \Delta$ it holds that
  \begin{equation*}
  \lspost_{\Delta'}^{*}(A) = (\mathcal{L}(\mathcal{T_{M}}) \cap
  \mathcal{L}(\mathcal{T}_{rh}) \cap \mathcal{L}(\mathcal{T}_{ht})
  \cap \mathcal{L}(\mathcal{T}_{ah}) \cap
  \mathcal{L}(\mathcal{T}_{cwf}) \cap \mathcal{T}^{-1}_{ct}(A) \cap
  \mathcal{L}(\mathcal{T}^{c}_{\Delta'}))|_{\cut}\ .
\end{equation*}

\end{proposition}

After assuring oneself that the stated tree automata calculate the
correct information, firstly one needs to understand that each
accepted tree is schedulable in the desired way.  Again we need to
consider only non-reentrant operations. In a first phase, after
reaching the intermediate configuration, the \emph{open} uses which
contain a cut are \emph{closed}.  This can be done based on a
topological ordering of the acyclic release graph, because for an open
use of lock $x$ only the predecessors of $x$ in the release graph
are needed to close the use, and because the tree automaton $\mathcal{T}_{ht}$
assures that none of these locks are finally acquired before the
cut. Again, the well-nestedness of lock operations allows to schedule
uses atomically. Afterwards, in a second phase, we schedule the rest
of the tree: We first schedule all uses, such that every thread is
either in its final state, or before a final acquisition. This is
possible, as none of the used locks has been finally acquired before
the cut.  Then, we schedule a final acquisition of a lock that is
minimal in the acquisition graph. As the acquisition graph is acyclic,
we always find such a lock. Moreover, the construction of the
acquisition graph guarantees that this minimal lock is not required to
schedule the rest of the tree. After the final acquisition, we
schedule all subsequent uses. We then continue with the next final
acquisition, until all threads have reached their final states.

Secondly, in order to show that the criteria checked by the tree
automata are necessary for a schedule to exist (again, reentrant
operations can be ignored), note that a schedulable tree obviously
contains no two final acquisitions of the same lock, nor a final
acquisition of a lock before the cut, that is used after the
cut. Moreover, the acquisition graph captures ordering constraints
between the final acquisitions and uses: An edge $x\to y$ means that
the final acquisition of $x$ must be scheduled before a use of
$y$. Also, all uses of $y$ must be scheduled before a potential final
acquisition of $y$. Thus, a cycle in the acquisition graph indicates
an unavoidable deadlock for any schedule of the tree.  Symmetrically,
an edge $x\to y$ in the release graph indicates that, after the
intermediate configuration has been reached, lock $x$ must be used
before an open use of $y$ can be closed. As an open use of a lock must
be closed before the lock can again be used, a cycle in the release
graph indicates an unavoidable deadlock after the intermediate
configuration has been reached.

\begin{example}
  An example application for this technique is solving forward bit
  vector problems bitwise.  Assume there is a set $G\subseteq \Gamma$
  of stack symbols corresponding to program points where the bit in
  question has just been generated. For simplicity we assume that
  generating the bit does not use any locks and does not spawn new
  threads.  The tree automaton $\mathcal{T}_{G}$ checks that some
  thread has just reached a top of stack symbol from $G$. The state
  space is $\mathbb{B}$, $\top$ is the accepting state and the rules
  are:
  \begin{align*}
    \nil_\gamma &\taarrow \top \quad \mif \gamma \in G&
    \nil_\gamma &\taarrow \bot \quad \mif \gamma \notin G\\
    \ret &\taarrow \bot & \cut\ p &\taarrow p\\
    \base\ p &\taarrow p &  \acq\ p &\taarrow p\\
    \calln\ p &\taarrow p &  \callr\ p\ q &\taarrow p \vee q\\
    \use\ p\ q &\taarrow p \vee q & \spawn\ p\ q &\taarrow p \vee q
  \end{align*}
  Here we use $\vee$ to denote the disjunction of the states
  interpreted as $true$ and $false$.  Let $\Delta' \subseteq\Delta$ be
  the set of transitions which do not kill the bit.  Then, the bit
  reaches a program point corresponding to a set of stack symbols $P$
  iff
  $\lspost^{*}_{\Delta'}(\lspost_{\Delta}^{*}(\{\nil^{p_{0}}_{\gamma_{0}}\})
  \cap \mathcal{L}(\mathcal{T}_{G}))\cap \mathcal{L}(\mathcal{T}_{P})
  = \emptyset$. Here $\mathcal{T}_{P}$ is defined analogously to
  $\mathcal{T}_{G}$ with $P$ instead of $G$. This allows, for
  instance, to calculate lock-sensitive def-use dependencies, which we
  use in our application.
\end{example}

Note that instead of projecting out the cut-nodes it is convenient to
leave them in the execution tree as they carry additional information
which can be useful if one wants to directly design a tree automaton
which checks for some property.  One then annotates the cut-nodes with
an index corresponding to their \emph{level}. This is done in our
actual implementation.


\section{Analysis of Java Programs}
\label{sec:java}

\subsection{Undecidability in the Presence of \code{wait}-Calls}
\label{sec:undec-pres-codew}

Monitor-DPNs assume that lock usages can be bound to procedure calls
which requires that the locks are used in a syntactically well-nested
fashion.  It seems that, by the very nature of block structure, locks
are used in a well-nested fashion in synchronized-blocks and -methods
in Java. However, this is no longer true in presence of calls to the
\code{wait}-method of an object used as a lock.  Once the
\code{wait}-method is called on a lock, the corresponding lock is
released, independently of whether it is the lock that has been
acquired last or not. This clearly breaks well-nestedness. Note that,
due to reentrance, well-nestedness can be broken even if the
\code{wait}-method is called only on the lock of the innermost
enclosing synchronized block or method, that is even if locking is
syntactically well-nested.  Ignoring this effect would render our
analysis unsound as it would consider too few schedules.

Unfortunately there is no hope to handle this construct completely in
the presence of recursion.  We show now that reachability is
undecidable in presence of \code{wait}-calls.  As done by Kahlon {\it
  et al.} \cite{DBLP:conf/cav/KahlonIG05} for general non-well-nested
locking it suffices to simulate pairwise rendezvous.  Based on this
one can reduce the undecidable emptiness problem for the intersection
of context free languages to a reachability problem as done by
Ramalingam \cite{Ramalingam:2000:CSA:349214.349241}.  We need three
auxiliary locks and one lock for each symbol to be communicated.
Initially, the first thread holds the locks $\{x_0, \dots x_n, t\}$
and the other $\{s, r\}$.  The following code models the synchronous
communication of the symbol associated with $x_i$:
\lstset{language=myjava}
\begin{center}
  \begin{minipage}{.8\linewidth}
\begin{lstlisting}
$x_i$.wait(1);synchronized($r$){};synchronized($s$){};$t$.wait(1);
synchronized($x_i$){};$r$.wait(1);$s$.wait(1);synchronized($t$){};
\end{lstlisting}
  \end{minipage}
\end{center}

These lines can only be executed in lock step as each thread relies on
the other thread offering its lock.  Each thread needs to hold two
locks so that no thread can step through its block multiple times
while the other offers a lock.  The initial configuration in which
both threads hold the appropriate locks can easily be enforced with
two additional locks and the following code at the beginning.  Here at
points $A$ and $B$ the actual code is inserted and the points directly
afterwards are checked for parallel reachability.  We use
\lstinline|sync($l_0\dots l_n$){S}| to abbreviate
\lstinline|synchronized($l_0$){$\dots$synchronized($l_n$){S}$\dots$}|.
\begin{center}
  \begin{minipage}{.73\linewidth}
\begin{lstlisting}
sync($a,x_0, \dots x_n, t$){synchronized(b){};a.wait(1); /*A*/}
sync($b,s,r$){b.wait(1);synchronized(a){}; /*B*/}
\end{lstlisting}
  \end{minipage}
\end{center}

\subsection{Implementation}
\label{sec:implementation}
We implemented our approach for the analysis of concurrent Java
programs. The analysis of a Java program consists of two phases.  In
the first phase a DPN-model of the program under analysis is generated
and in the second phase this model is analysed using the described
techniques. 

\subsubsection{Model Generation}
\label{sec:model-generation}
For model generation we utilize the T.J. Watson Libraries for Analysis
(WALA)\footnote{\url{http://wala.sf.net}} targeting Java-Bytecode.
WALA provides control flow and points-to information in the form of
control flow, call and heap graphs.  Following Esparza and Knoop
\cite{Esparza:1999}, we encoded the local state in the stack symbols
of the DPN, which consist, in the most basic setting, only of the
control point of the program.  The control states of the DPN are used
to model the thread state (e.g.\ thrown exceptions) and return
information of synchronized blocks as these are modeled as multi-exit
procedures.  Modeling exceptions precisely is important as they can be
thrown by most Java-instructions.  If one would abstract from
exceptions (modeling them as regular control flow) the DPN could for
instance skip uses of locks in called procedures by using abstracted
exceptional control flow.  By explicitly modeling the exceptional
flow, we avoid that in the DPN critical parts are reachable spuriously
after using exceptional control flow.

In order to model lock operations in the DPN, in a first step a finite
set of possible locks has to be identified.  As locks are bound to
dynamically allocated objects in Java, we identify statically unique
objects in the points-to information provided by WALA with a simple
reachability analysis on the call and control flow graphs.  That is,
we search for abstract objects for which we can guarantee that there
will be at most one instance at runtime because either the associated
allocation instruction can be executed at most once, or they represent
a class object.  Moreover, we implemented a version of the random
isolation technique from Kidd {\it et al.} \cite{random-isolation}.
This technique \emph{randomly isolates} one instance of a class of
objects and by proving the desired property for the isolated instance
concludes to the validity of the property for all instances. This is
possible if the property to be checked is, in some sense, coupled to
an instance, for example, a data race or def/use dependency that
occurs on a field of a specific instance.  This is useful if, for
instance, access to a field is guarded by the object it belongs to.
Then, one can assume that either two threads use the same lock object,
or they will access different fields.  While Kidd {\it et al.}
implemented random isolation as a source code transformation, we
integrated this technique directly into the DPN-model, which can
easily keep track of the needed information in its local state.  As
already mentioned, one needs to take care of potential
\code{wait}-calls in the program.  We use a simple data-flow analysis
to statically identify monitors inside which such a call may occur and
approximate these as regular procedure calls in the DPN-model.

The DPN-model generated in this way is in many cases too big to be
handled directly.  For example a single call to some Java API methods
can generate more than 100\,000 control flow instructions.  We
therefore use a simple pruning technique on the call graph to reduce
the size of the model by automatically removing \emph{uninteresting}
procedure calls and overapproximating their possible effects on the
control state.  This technique depends on the analysis instance and
therefore different models have to be generated for different
instances. E.g., for a data race analysis methods which do
neither directly nor indirectly use any locks or access the field in
question are uninteresting and can be pruned.  Most analyses like bit
vector analyses posses a similar property which allows to prune the
model accordingly.  Further pruning on the instruction level would
also be possible, but was not implemented yet.

\subsubsection{DPN Analysis}
\label{sec:dpn-analysis}
For analysis we translate the generated DPN-model to a tree automaton
which is encoded in a logical program for the XSB
system\footnote{\url{http://xsb.sf.net}}.  XSB is a logic programming
and deductive database system.  This allows for a simple symbolic
encoding of the automata manipulating, for instance, sets of locks or
graphs.  The used tree automata operations can be implemented
straightforwardly in a logical program utilizing the supported tabled
evaluation.  To alleviate the exponential blow up in the number of
locks, we utilize a special emptiness check for product tree automata,
which explores one component top-down and the other bottom-up.  If one
inspects the defined automata, one recognizes that some of them are
extremely nondeterministic when evaluated upwards and others when
evaluated downwards, but \emph{nearly deterministic} in the other
direction.  The emptiness check is implemented by a predicate which
marks states for which there exists a tree recognized with them.  By
ordering the premises of the non-emptiness predicate suitably, the
system evaluates the automata appropriately in different directions.
As another optimization, we use XSB's answer subsumption based upon a
partial order on the states of the automata.  That is, if we have two
states with the same \emph{conflict potential} and one state has a
strictly more restrictive acquisition/release history, then that state
can be dropped.  By instrumenting the non-emptiness predicate, a
witness for the execution violating the checked property can be
constructed.

\subsection{Application}
\label{sec:empirical-results}
As an application for the simple reachability analyses we developed a
plugin for the Eclipse IDE that runs data race checks on shared fields
as described in \autoref{ex:data-race}.  Checking small programs (up
to 1k LOC) takes between several milliseconds up to a few seconds (not
checking library fields).  We investigated several found races
manually based on the generated witnesses.  Many were actual races,
mostly found in exception handlers.  False positives were often due to
the inability to model the used locks in the DPN and the imprecise
field abstraction based on points-to information.

As a second application, the iterated reachability analyses were
integrated as a def-use dependency checker into the Joana tool for
information flow control\footnote{\url{http://joana.ipd.kit.edu}}.
Joana builds a system dependency graph (SDG) that contains so called
\emph{interference edges}, which model def-use dependencies between
different threads.  After generating the system dependency graph with
Joana, we check individual interference edges for feasibility.  That
is, we check whether the writing instruction can be executed before
the reading instruction without an intervening \emph{killing}
instruction.  Recall the examples from \autoref{tab:ex1} in
\autoref{sec:introduction}, where, in an information flow control
setting, we want to prove that there is no information flow from the
assignment of 42 to the \code{print} statement.  Effects of thread
creation like in the first example are already handled by Joana, hence
it is able to prove the absence of information flow in the first
example.  In the other examples Joana suspects an information flow
which is due to spurious interference edges in the second, third,
fifth and sixth example.  Our analysis is able to remove these
spurious interference edges in these examples, allowing Joana to prove
the absence of information flow.  In the fourth example both
interference edges are feasible on their own, but not both within one
execution.  We did not yet implement automatic handling of this
effect, but were able to handle this with our analysis in a mock-up
setting.  We intend to integrate handling of those effects with a
deeper integration into the slicing algorithm of Joana.  This will
need to check only \emph{critical} SDG-subpaths for feasibility, as
otherwise there are too many paths to be checked.

In order to test performance of the analysis for real world
applications, we also tried to eliminate interference edges assumed by
Joana in a suite of 24 programs.  The suite contained concurrent
programs from different benchmark suites and some genuine JavaME
programs.  The programs contained a total of 32\,310 lines of code
(excluding library code) ranging from 29 to 6\,252 LOC with an average
of 1\,346.  In total, 10\,376 dependencies have been checked.
Analysis time per dependency varied between 0.16 and 549 seconds with
an average of 12 seconds.  The number of transitions of the DPN-models
ranged from 140 to 8\,801 with an average of 3\,393.  We could not
find additional infeasible dependencies within these applications.
This is likely due to the fact that Joana already excludes many
dependencies which are infeasible due to effects of thread creation
and only twelve of the programs contained abstractable locks (with a
maximum of two). A further reason might be that, due to possibly
spurious exceptional control flow, some of the more subtle effects
treated by our analysis are not visible in the abstract program model.


\section{Conclusion}
\label{sec:conclusion}
The aim of this work was to implement DPN-based reachability analyses
and explore their applicability for the improvement of an information
flow control analysis for Java.  For this different existing and new
techniques had to be combined on a level appropriate for
implementation.  Notably, we introduced an iterable, tree-regular
characterization of reachable configurations for DPN, based on
execution trees.  This allows us to conclude that post$^{*}$ preserves
tree-regularity.  We gave a modular description of the relevant
constructions in the form of finite tree automata.  This was in
particular critical for obtaining an efficient evaluation strategy.
The implementation of the non-iterated analysis performs well in most
cases and yields encouraging results.  The nondeterministic placement
of the cut impacts on the performance of the iterated analysis and
encourages the development of further optimizations.

\paragraph{Acknowledgments}
\label{sec:acknowledgments} 

Using DPN analysis in the context of IFC-Analysis for Java was
envisioned in a joint DFG project with Gregor Snelting's group from
Karlsruhe Institute of Technology (KIT) who develop the Joana
framework. We thank our colleagues from KIT for the good
collaboration, in particular Jürgen Graf who helped us getting started
with the WALA framework and integrating the analysis into the Joana
framework and Dennis Giffhorn who assembled the test suite that was
used. We thank Alexander Wenner for lots of helpful discussions
especially on the decidability issue in the presence of
\code{wait}-calls.

\medskip\noindent \emph{We dedicate this paper to Dave Schmidt, one of
  the pioneers in studying connections between model checking and data
  flow analysis, at the occasion of his 60th birthday.  }


\bibliography{main}

\end{document}